\documentclass[a4paper,12pt]{article}
\usepackage{jheppub,esint,shuffle,psfrag}

\usepackage{esint} 
\usepackage{breqn}
\usepackage{amsmath,bm}
\usepackage{hyperref}
\usepackage{mathrsfs} 

\definecolor{myred}{RGB}{233, 33, 45}

\makeatletter
\DeclareFontFamily{OMX}{MnSymbolE}{}
\DeclareSymbolFont{MnLargeSymbols}{OMX}{MnSymbolE}{m}{n}
\SetSymbolFont{MnLargeSymbols}{bold}{OMX}{MnSymbolE}{b}{n}
\DeclareFontShape{OMX}{MnSymbolE}{m}{n}{
    <-6>  MnSymbolE5
   <6-7>  MnSymbolE6
   <7-8>  MnSymbolE7
   <8-9>  MnSymbolE8
   <9-10> MnSymbolE9
  <10-12> MnSymbolE10
  <12->   MnSymbolE12
}{}
\DeclareFontShape{OMX}{MnSymbolE}{b}{n}{
    <-6>  MnSymbolE-Bold5
   <6-7>  MnSymbolE-Bold6
   <7-8>  MnSymbolE-Bold7
   <8-9>  MnSymbolE-Bold8
   <9-10> MnSymbolE-Bold9
  <10-12> MnSymbolE-Bold10
  <12->   MnSymbolE-Bold12
}{}

\let\llangle\@undefined
\let\rrangle\@undefined
\DeclareMathDelimiter{\llangle}{\mathopen}%
                     {MnLargeSymbols}{'164}{MnLargeSymbols}{'164}
\DeclareMathDelimiter{\rrangle}{\mathclose}%
                     {MnLargeSymbols}{'171}{MnLargeSymbols}{'171}
\makeatother

\newcommand{\bs}{\begin{shaded}}
\newcommand{\es}{\end{shaded}\noindent}
\def\ba#1\ea{\begin{align}#1\end{align}}		
\newcommand{\be}{\begin{equation}}
\newcommand{\ee}{\end{equation}}
\newcommand{\mc}{\mathcal }

\newcommand{\la}{\label}

\DeclareMathOperator{\tr}{\text{tr}}

\newcommand\re[1]{(\ref{#1})}

\newcommand\lr[1]{{\left({#1}\right)}}

\newcommand \VEV [1] {\Big\langle{#1}\Big\rangle}
\newcommand \ket [1] {|{#1}\rangle}
\newcommand \bra [1] {\langle {#1}|}

\def \qqqquad {\qquad\qquad}

\newcommand{\ft}[2]{{\textstyle\frac{#1}{#2}}}

\def\numberbysection{\@addtoreset{equation}{section}
                     \def\theequation{\thesection.\arabic{equation}}}

\newcommand{\ci}{\cite}

\def\OO{\mc O}

\def \iffa {\iffalse}

\newcommand{\sql}{\sqrt\l}
\renewcommand{\l}{\lambda}

\newcommand{\vev}[1]{\big\langle #1 \big\rangle}
\newcommand{\vvev}[1]{\big\llangle #1 \big\rrangle}

\newcommand{\QQ}{\mathsf{Q}_{L}}

\begin{document}

\vspace{-4cm }

\begin{flushleft}
 \hfill \parbox[c]{30mm}{
 IPhT--T23/113 }
\end{flushleft}
\author{M. Beccaria$^a$ and G.P. Korchemsky$^{b}$}
\affiliation{
$\null$
$^a$Universit\`a del Salento, Dipartimento di Matematica e Fisica \textit{Ennio De Giorgi},\\ 
\phantom{a} and I.N.F.N. - sezione di Lecce, Via Arnesano, I-73100 Lecce, Italy
\\		
$\null$
$^b${Institut de Physique Th\'eorique\footnote{Unit\'e Mixte de Recherche 3681 du CNRS}, Universit\'e Paris Saclay, CNRS,\\ \phantom{a}  91191 Gif-sur-Yvette, France}   
}
\title{Four-dimensional $\mathcal N=2$ superconformal long circular quivers}
\abstract{
\small
We study four-dimensional $\mathcal N=2$ superconformal circular, cyclic symmetric quiver theories which are planar equivalent to $\mathcal N=4$ super Yang-Mills. 
We use localization to compute nonplanar corrections to the free energy and the circular half-BPS Wilson loop in these theories for an arbitrary number of nodes, and examine their behaviour in the limit of long quivers. Exploiting the relationship between the localization quiver matrix integrals and  an integrable Bessel operator, we find a closed-form expression for the leading nonplanar correction to both observables in the limit when  the number of nodes and 't Hooft coupling become large. We demonstrate that it has different asymptotic behaviour depending on how the two parameters are compared, and interpret this behaviour in terms of properties of a lattice model defined on the quiver diagram. }

\maketitle
 
\section{Introduction and summary}

In this paper, we continue the study  initiated in \cite{Beccaria:2023kbl} of a special class of four-dimensional $\mathcal N=2$ superconformal gauge theories which are planar equivalent to $\mathcal N=4$ SYM theory. A distinguished feature of these theories is that various observables, e.g. free energy on four-sphere and circular half-BPS Wilson loop, can be computed exactly, for arbitrary value of the coupling constant and rank of the gauge group, in terms of matrix model integrals using the localization technique \cite{Pestun:2009nn,Pestun:2016zxk}.  
The actual evaluation of these matrix integrals turns out to be a nontrivial task due to a complicated non-polynomial
form of interaction potential. The latter is given by an infinite sum of single and double trace terms \cite{Pini:2017ouj,Billo:2019fbi,Fiol:2020bhf,Fiol:2020ojn}. 

In the previous work \cite{Beccaria:2023kbl}, we developed a technique for the systematic expansion of such matrix integrals at large $N$ and applied it to determine the nonplanar corrections in various $\mathcal N=2$ superconformal theories including the two-nodes quiver theory with the gauge group $SU(N)\times SU(N)$. Here we shall apply this technique to compute the free energy and circular half-BPS Wilson loop in $\mathcal N=2$ superconformal circular quiver theory  for an arbitrary number of nodes $L\ge 2$. In what follows we refer to this theory as $\mathsf Q_L$ model.

\begin{figure}[t!]
\begin{center}
\psfrag{L}[cc][cc]{$L$}\psfrag{1}[cc][cc]{$1$}\psfrag{2}[cc][cc]{$2$}\psfrag{3}[cc][cc]{$3$}
\includegraphics[width=0.3\textwidth]{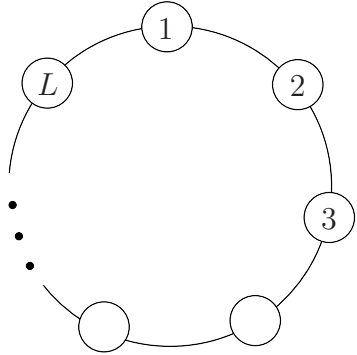}
\caption{Diagrammatic representation of the circular quiver theory. Each node represents $SU(N)$ vector multiplet and lines connecting neighbouring nodes represent hypermultiplets in the bi-fundamental representation of $SU(N)\times  SU(N)$. In the localization matrix model representation, the same diagram defines a lattice model with $L$ sites equipped with a nearest neighbour  interaction.}
\label{fig:qiuv}
\end{center}
\end{figure}

The field content of the $\mathsf Q_L$ model can be represented as a circular quiver diagram shown in Figure~\ref{fig:qiuv}.
This model has the $SU(N)^{L}$ gauge symmetry and 
depends in general on $L$ independent 't Hooft couplings, one per each $SU(N)$ factor. At strong coupling, it is dual to type IIB string theory on the AdS${}_5\times (S^5/\mathbb Z_L)$ background \cite{Kachru:1998ys}. The planar limit of the $\mathsf Q_L$ model was studied both at weak and strong  't Hooft couplings in \cite{Mitev:2014yba,Mitev:2015oty,Fiol:2015mrp,Fiol:2020ojn,Zarembo:2020tpf,Ouyang:2020hwd,Billo:2021rdb,Galvagno:2020cgq,Beccaria:2021ksw,Pini:2017ouj,Billo:2022gmq,Billo:2022fnb,Preti:2022inu}.
For simplicity, we will assume that all couplings are equal and denote them as $\lambda$. In this case, the  $\mathsf Q_L$ model coincides with the $\mathbb Z_L$ orbifold of $\mathcal N=4$ SYM with the $SU(L N)$ gauge group \cite{Lawrence:1998ja}. 

In the planar limit, the  $\mathsf Q_L$ model  is equivalent to $L$ copies of $\mathcal N=4$ SYM with the $SU(N)$ gauge group but the two models are different beyond this limit. 
In particular, the free energy of this model defined on the unit four-sphere has the following form
\begin{align}\label{free}
F_{\QQ}=L F_{\mathcal N=4}+\Delta F_{\QQ}\,,
\end{align}
where $F_{\mathcal N=4}=-\frac12(N^2-1)\log \lambda$ is the free energy of $\mathcal N=4$ SYM theory with the $SU(N)$ gauge group. The nonplanar correction to the free energy
 $\Delta F_{\QQ}$  admits an expansion in powers of $1/N^2$
\begin{align}\label{F-exp}
\Delta F_{\QQ} = F_L^{(0)} + {1\over N^2}  F_L^{(1)} + O(1/N^4)\,.
\end{align}
Our goal in this paper is to determine the dependence of the coefficient functions  $F_L^{(0)}, F_L^{(1)}, \dots$ on 't Hooft coupling $\lambda$ and the number of nodes $L$.  
For the simplest, two-nodes quiver $\mathsf Q_{L=2}$ model,  this problem was studied in  \cite{Beccaria:2023kbl}.

Another motivation of the present work  is to explore the limit of large number of nodes in the $\mathsf Q_L$ model. 
This limit has been previously discussed in application to superconformal theories with eight supercharges that correspond to conformal fixed points of linear quiver gauge theories in $d$ dimensions \cite{Uhlemann:2019ypp,Uhlemann:2019lge,Coccia:2020cku,Coccia:2020wtk,Akhond:2022oaf}. In the special case of $d=4$ relevant to our discussion, four-dimensional $\mathcal N=2$ superconformal long linear quiver theories were recently analyzed at large $N$ in \cite{Nunez:2023loo}. Unlike the $\mathsf Q_L$ model, these theories differ from $\mathcal N=4$ SYM already in the planar limit.
The leading $O(N^2)$ contribution to the free energy in these theories is a nontrivial function of the number of nodes $L$ and 't Hooft coupling. At strong coupling and large $L$, its asymptotic behaviour depends on the order of limits -- the free energy of the long linear quiver grows as $O(N^2 L)$ if the number of nodes is much larger than the 't Hooft coupling and  $O(N^2 L\log L)$ otherwise. In contrast, the free energy of the circular quiver theory \re{free} grows as $F_{\QQ}=O(N^2L)$ at large $L$, independently on the value
 of the 't Hooft coupling constant. We compute below the leading nonplanar correction to the free energy \re{free} and show that, similar to the planar 
 contribution to the free energy of the long linear quiver, its asymptotic behaviour at strong coupling and large $L$ depends on how these two parameters are compared (see Eqs.~\re{f-fun} and \re{log-l} below).
   
The localization yields the partition function of the $\QQ-$model on the unit sphere, $Z_{\QQ}=e^{-F_{\QQ}}$, as an integral over the $SU(N)$ matrices $A_I$ describing zero modes of scalar fields 
in vector multiplets at all nodes $I=1,\dots,L$ (see \re{Z-repr} below) \cite{Pestun:2009nn,Pestun:2016zxk}. 
Discussing its dependence 
on the number of nodes, it is advantageous to think about the quiver diagram shown in Figure~\ref{fig:qiuv} as defining a one-dimensional lattice model with $L$ sites. The degrees of freedom at each site are described by the matrices $A_I$  subject to periodic boundary condition $A_{I+L}=A_I$. They interact among themselves as well as with their nearest neighbours at the sites $I\pm 1$. 
The form of the interaction potential is fixed by the potential of the localization matrix model. The free energy of the $\QQ-$model  \re{free}  coincides with the free energy of this lattice model. 

Using this identification, the free energy $F_{\QQ}$ can be expressed as the sum 
over excitations propagating across the one-dimensional lattice shown in Figure~\ref{fig:qiuv} and interacting with each other.  The number of sites can be arbitrary $2\le L<\infty$, but the limit of large $L$ is of a special interest. In this limit, the lattice degenerates into a circle with circumference $O(L)$ and the free energy of the lattice model is expected to exhibit an extensive behaviour, $F_{\QQ}=O(L)$, provided that the effective interaction between different sites is short-range. 
The last condition depends on the values of $1/N$ and $\lambda$. 

As mentioned above, the planar contribution to the free energy \re{free} grows linearly with $L$ for an arbitrary coupling constant $\lambda$. The question arises whether the same property holds for 
the nonplanar correction to the free energy \re{free} 
\begin{align}\label{density}
\Delta F_{\QQ} \stackrel{?}{=}  L\, \varepsilon(\lambda,1/N) + O(L^0)\,.
\end{align}
Here the energy density $\varepsilon(\lambda,1/N)$ depends on 't Hooft coupling and has large $N$ expansion similar to \re{F-exp}. We show below that the relation \re{density} holds at weak coupling.  At strong coupling,  the validity of \re{density} depends on how $L$ compares to $\sqrt\lambda$. 
We find that the relation \re{density}  is verified for $L\gg\sqrt\lambda\gg 1$. In the opposite limit, for $ \sqrt\lambda\gg L\gg 1$, the free energy $\Delta F_{\QQ}$ receives corrections that run in powers of $L/\sqrt\lambda$ and, therefore, violate \re{density}. In this case, the lattice model becomes strongly correlated and the above mentioned condition is not satisfied. 

To verify the relation \re{density}, we computed the first two terms of the large $N$ expansion of the free energy \re{F-exp} for arbitrary number of nodes $L$. We show below that, for arbitrary coupling $\lambda$, they can be expressed in terms of a semi-infinite matrix whose entries are the coefficients of the double trace terms entering the potential of the localization matrix model. The same entries can be identified
as
matrix elements of a certain integral operator known in the mathematical literature as a truncated Bessel operator.~\footnote{It is interesting to notice that the same operator previously appeared in the study of level spacing distribution in the matrix models \cite{Tracy:1993xj} and, more recently, in the context of the AdS/CFT correspondence \cite{Belitsky:2020qrm,Belitsky:2020qir}.} 
Applying the technique of \cite{Beccaria:2023kbl} and exploiting the known properties of the Bessel operator, we computed $F_L^{(0)}$ and $F_L^{(1)}$ for arbitrary number of nodes $L$ both at weak and strong coupling.

At weak coupling, the free energy $\Delta F_{\QQ}$ is given by series in  't Hooft coupling that starts at order $O(\lambda^2)$. The expansion coefficients are given by multilinear combinations of odd Riemann zeta values $\zeta(2n+1)$ accompanied by rational $L-$dependent coefficients (see \re{F-weak} and \re{F1-weak} below). Going to the limit of large $L$, we find that, in agreement with \re{density}, the free energy $\Delta F_{\QQ}$ grows linearly with the number of nodes  
\begin{align} \label{lin-weak} 
\Delta F_{\QQ} = L \,\left[\frac{3\zeta(3)}{128\pi^{4}}\,\l^{2}-\frac{15\zeta(5)}{1024\pi^{6}}\,\l^{3} +O(
\l^4) +O(1/N^2)\right]  \,.
\end{align}
This behaviour is not surprising because the effective interaction between different sites of the quiver diagram in Figure~\ref{fig:qiuv} is short range at weak coupling and, as a consequence, the free energy has an extensive behaviour. 

At strong coupling, the first two terms of the expansion of the free energy \re{F-exp} are given by
 \begin{align}\notag\label{int:res1} 
 F_L^{(0)}{}&=\sqrt\lambda\frac{ \left(L^2-1\right)}{6 L}  -\frac12  (L-1) \log \lambda + O(\lambda^0)\,,
\\[2mm] 
F_L^{(1)}{}&=-\lambda^{3/2} \frac{(L^2-1) (L^2+1)}{1920 L^3}+\lambda \frac{(L^2-1) (L^2-9)}{5760 L^3} +O(\sqrt\lambda)\,.
\end{align}
The subleading corrections to both relations run in powers of $1/\sqrt\lambda$, their explicit expressions can be found in \re{FL-free} and \re{F1-res} below.
As expected, the functions \re{int:res1} have a nontrivial dependence on the number of nodes $2\le L<\infty$. 
For $L=2$, the relations \re{int:res1} reproduce the analogous expressions for the free energy in the $\mathsf{Q}_{2}$ model obtained in \cite{Beccaria:2023kbl}. 
Notice that the functions \re{int:res1} vanish for an unphysical value $L=1$. In this case, the quiver diagram in Figure~\ref{fig:qiuv} contains only one node.  The corresponding localization matrix integral for $F_{\QQ}$ becomes Gaussian and it coincides with the free energy of $\mathcal N=4$ SYM. As a consequence, for $L=1$ the nonplanar correction $\Delta F_{\QQ}$ has to vanish for arbitrary coupling constant.

At strong coupling, we can use \re{int:res1} to verify that for $L\gg 1$ the first few terms of the strong coupling expansion of  the free energy \re{F-exp} grow linearly with $L$. However the situation becomes more complicated when we include the subleading corrections in $1/\sqrt\lambda$. 
They take the following form at large $L$
\begin{align} \label{f-fun} 
\Delta F_{\QQ} {}& = L \left[  {1\over 6} \sqrt\lambda -\log {\sqrt\lambda\over 4\pi}  + \lr{\tfrac14-6\log\mathsf A}  -\frac{1}{4\sqrt\lambda} \log L\right]+ f\lr{L\over \sqrt\lambda}+\dots \,,
 \end{align}
where $\mathsf A$ is the Glaisher constant and dots denote corrections suppressed by powers of $1/N^2$ and $1/L$.
At large $L$ and $\sqrt\lambda$, the subleading corrections to \re{f-fun} run in powers of $L/ \sqrt\lambda$. 
They are described by the function $f(L/ \sqrt\lambda)$. 

Remarkably, this function can be found in a closed form (see \re{f-resum} below). 
It is interesting to examine its behaviour for different values of the ratio $l=L/\sqrt\lambda$.
At small  $l$,  or equivalently $L \ll \sqrt\lambda$, the function $f(l)$  is given by a series in $l$ that starts at order $O(l^2)$.
The resulting expression for the free energy \re{f-fun} contains  $(L\log L)/\sqrt\lambda$ term and does not satisty \re{density}. The reason why the free energy \re{f-fun} does not grow linearly with $L$ is that the quiver lattice model becomes strongly correlated for $\sqrt\lambda \gg L \gg 1$, thus invalidating the relation \re{density}. 

At large $l$, or equivalently $L\gg \sqrt\lambda$, the function $f(l)$ behaves as $f(l) =  l (\log (4\pi l)-c)/4 + O(l^0)$, where $c$ is a constant defined in \re{f-as} below. Substituting this relation into \re{f-fun} we find that the term proportional to $(L\log L)/\sqrt\lambda$ cancels and we recover the expected scaling behaviour \re{density}
\begin{align}\label{log-l}  
\Delta F_{\QQ} {}& = L \left[  {1\over 6} \sqrt\lambda - \log {\sqrt\lambda\over 4\pi} + \lr{\tfrac14-6\log\mathsf A}  -\frac{1}{4\sqrt\lambda} \lr{ \log {\sqrt\lambda\over 4\pi}+c} \right] +O(1/N^2)+O(L^0) \,.
 \end{align} 
We would like to emphasize that this relation holds for $L\gg \sqrt\lambda\gg 1$. As compared with \re{f-fun}, the term proportional to $\log L$ gets replaced in \re{log-l} with $\log (\sqrt\lambda/(4\pi))+c$.

Another interesting quantity that can be computed in the $\mathsf Q_L$ model using the localization technique is the expectation value of the half-BPS circular Wilson loop defined at one of the nodes of the quiver. In virtue of planar equivalence of the $\mathsf Q_L$ model and $\mathcal N=4$ SYM, the  expectation value of the Wilson loops  in these two theories coincide up to nonplanar corrections. We find that the leading nonplanar correction to their ratio it is proportional to a derivative of the free energy
\begin{align}\label{rW} 
{W_{\QQ}\over W_{\mathcal N=4}} = 1 - \frac{1}{4 L N^2}\lambda^2\partial_\lambda \Delta F_{\QQ}+ O(1/N^4)\,.
\end{align}
This relation is valid for arbitrary coupling $\lambda$ and the number of nodes $L$.   Similar relation has been previously derived for other superconformal $\mathcal N=2$ theories   including the $\mathsf Q_2$ model \cite{Beccaria:2021ism,Beccaria:2023kbl}.  
 
At large $L$, the scaling behaviour \re{density} of the free energy implies that the nonplanar $O(1/N^2)$ correction to \re{rW} is independent of the number of nodes $L$. As explained above, this property holds both  at weak coupling and at strong coupling for $1\ll \sqrt\lambda \ll L$. In the latter case, we have
\begin{align}\label{W-lim}
{W_{\QQ}\over W_{\mathcal N=4}} {}& =1 -\frac{1}{N^2}\left[\frac{\lambda ^{3/2}}{48}-\frac{\lambda }{8}+\frac{\sqrt{\lambda } }{32} \lr{ \log {\sqrt\lambda\over 4\pi}+c-1}
+ O(\lambda^0)\right]+O(1/N^4)+O(1/L)\,.
\end{align}
 At very strong coupling, for $1\ll L \ll \sqrt\lambda$, we find instead
\begin{align}\label{W-lim1}
{W_{\QQ}\over W_{\mathcal N=4}} {}& =1 -\frac{1}{N^2}\left[\frac{\lambda ^{3/2}}{48}-\frac{\lambda }{8}+\frac{\sqrt{\lambda } }{32}  \log L
+ O(\lambda^0)\right]+O(1/N^4)+O(1/L)\,.
\end{align}
Similar to the free energy,   $\log L$ term inside the brackets is replaced with $\log (\sqrt\lambda/(4\pi))+c-1$ for $\sqrt\lambda \ll L$.

It is interesting to note that the order of limit phenomenon that we described above for the free energy and the circular Wilson loop in the long quiver limit is not a specific feature of $\mathcal N=2$ superconformal theories. Analogous phenomenon has been previously observed in the study of four-point correlation functions of infinitely heavy half-BPS operators in planar $\mathcal N=4$ SYM in the so-called null limit when the four operators are light-like separated in a sequential manner, $(x_i-x_{i+1})^2\sim e^{-y}$ for $y\to \infty$ \cite{Coronado:2018cxj}. It turns out that, at strong coupling, these functions  have different asymptotic behaviour for $1\ll y \ll \sqrt\lambda$ and $1\ll  \sqrt\lambda\ll y$ \cite{Belitsky:2019fan,Bargheer:2019exp,Belitsky:2020qrm}.  

The limit of large $L$ can be viewed as a continuum limit of the lattice model defined on the quiver diagram shown in Figure~\ref{fig:qiuv}. The fact that the $L-$dependence of the free energy \re{f-fun}  
is encoded in a  function of the ratio $L/\sql$ suggests that interaction between excitations is characterized in this limit by  a correlation length $\xi =O(\sql)$. In application to the long circular quiver, we therefore expect that the correlation function of local (chiral primary) operators $O_{n,I}(x)=\tr(\varphi_I^n(x))$ placed at different nodes of the quiver  $|I-J|=O(L)$ has to scale as
$\vev{O_{n,I}(x) \bar O_{n,J}(0)}\sim e^{-|I-J|/\xi}$.

It would be interesting to reproduce the strong coupling expansion of the free energy of the long circular quiver theory \re{f-fun} using the AdS/CFT correspondence. In the holographic description, the $\QQ$ theory is dual to type IIB superstrings propagating on the orbifold AdS${}_5\times (S^5/\mathbb Z_L)$. To obtain the free energy, one has to determine higher-derivative string corrections to the type IIB 10d effective action.
For the simplest, length $L=2$ quiver, this problem was discussed in \cite{Beccaria:2023kbl}. The remarkable simplicity of the obtained expression \re{f-fun} (see also \re{f-resum}) suggests that the problem can be solved in the limit of long quivers. Another interesting question is to clarify the origin of the ratio $L/\sql$ on the string side.

The rest of the paper is organized as follows. In Section~\ref{sect2} we present the matrix model representation of the partition function of the $\mathsf Q_L$ model on the unit four-sphere. We then use it to derive a Feynman diagram representation of the first few terms in the large $N$ expansion of the free energy \re{F-exp}. The contribution of these diagrams to the free energy is evaluated in Section~\ref{sect3}. We show that it can be expressed in a concise way in terms of matrix elements of the resolvent of the so-called Bessel kernel. We exploit this relation to derive the free energy $F_{\QQ}$ at weak and strong coupling. In Section~\ref{sect:large} we examine behaviour of the free energy of  the $\mathsf Q_L$ model in the limit of large number of nodes $L$. The expectation value of half-BPS circular Wilson loop is computed in Section~\ref{sect5}. Some technical details are presented in two appendices. 
 
\section{Matrix model representation}\label{sect2}

Using the localization the partition function of the quiver $\QQ$ model defined on the unit sphere $S^4$ (with equal coupling constants on all nodes) can be 
expressed as a matrix integral~\cite{Pestun:2009nn,Pestun:2016zxk}
\begin{align}\label{Z-Q}
Z_{\QQ}  =\int \prod_{I=1}^L \bigg[ \prod_{r=1}^{N}da_{I,r} \,\delta\Big(\sum_{r}a_{I,r}\Big) \, \Delta^2(\bm{a}_I)\bigg]  \,e^{-S_{\QQ} (\bm{a}_1,\dots,\bm{a}_L)}\,,
\end{align}
where integration goes over eigenvalues $\bm{a}_I=\{a_{I,1},\dots,a_{I,N}\}$ of hermitian traceless $N\times N$ matrices $A_I$ describing zero modes of a scalar field in the vector multiplet on $S^4$ at nodes $I=1,\dots,L$. Here $\Delta(\bm{a}_I) = \prod_{r<s}(a_{I,r}-a_{I,s})$ is a Vandermonde determinant 
depending on the eigenvalues at the node $I$. 

The potential  in \re{Z-Q} is given by the sum of one-loop perturbative and instanton contributions. The latter can be neglected at large $N$ leading to  
\begin{align} \label{S-Q2}
S_{\QQ}  {}& =\sum_{I=1}^L \left[   \frac{8\pi^{2}N}{\l}  \sum_{r=1}^N  a_{I,r}^{2} 
+ \sum_{r,s=1}^N \Big(\log H(a_{I,r}-a_{I+1,s})-\log H(a_{I,r}-a_{I,s})\Big)\right],
\end{align}
where the periodicity condition $a_{I+L,r}=a_{I,r}$ is implied. This relation involves the $H-$function   given by the product of the Barnes $G-$functions
\ba\notag
H(x) & = e^{-(1+\gamma_{\rm E})\,x^{2}}
\,G(1+ix)G(1-ix)
\\
{}&
= \exp\left( \sum_{n=1}^{\infty}\frac{(-1)^{n}}{n+1}\zeta(2n+1)\,x^{2n+2}\right),
\la{23} 
\ea
where the second relation holds at small $x$ and it involves odd Riemann zeta values.
  
It is convenient to express the potential  (\ref{S-Q2}) in terms of traces of the hermitian matrices
\begin{align}\label{O}
\OO_{i}(A_I)= \tr \Big({A_I\over  \sqrt N}\Big)^{i} = \sum_{r=1}^N \left(a_{I,r}\over \sqrt N\right)^{i}\,,
\end{align}
where $i\ge 2$. Expanding the $H-$functions in (\ref{S-Q2}) in powers of eigenvalues $a_r$ and rescaling them as $a_r\to (8\pi^2 N/\lambda)^{-1/2} a_r$, we get
\begin{align} \la{24} 
S_{\QQ} {}&= \sum_{I=1}^L\left[ \tr A_I^2    - S_{\rm int}(A_I,A_{I+1})\right]\,,
\end{align}
where $A_{L+1}\equiv A_1$ and the interaction term is given by infinite bilinear combinations of the single traces (\ref{O})  
\begin{align}\label{Sint}
S_{\rm int}(A_I,A_{I+1})&= \frac18 \sum_{i,j\ge 2}C_{ij}(\lambda)  \left[ \OO_{i}(A_I)-\OO_{i}(A_{I+1}) \right]\left[ \OO_{j}(A_I)-\OO_{j}(A_{I+1}) \right].
\end{align}
The sum does not involve $\OO_{1}(A_I)=\tr(A_I/\sqrt N)$ because it vanishes for the $SU(N)$ matrices $A_I$.
The expansion coefficients   in \re{Sint} are different from zero only if the indices $i$ and $j$ have the same parity. Nonzero coefficients 
$C^+_{ij} (\lambda)\equiv C_{2i,2j}$ and $C^-_{ij} (\lambda)\equiv C_{2i+1,2j+1}$ are given by~\cite{Billo:2021rdb,Billo:2022fnb}
 \begin{align}\label{C}\notag
C^-_{ij} (\lambda)&= 8\,\Big(\frac{\l}{8\pi^{2}}\Big)^{i+j+1}\,(-1)^{i-j+1}\,\zeta(2(i+j)+1)\,\frac{\Gamma(2(i+j)+2)}{\Gamma(2i+2)\Gamma(2j+2)}\,,
\\
C^+_{ij} (\lambda)&= 8\,\Big(\frac{\l}{8\pi^{2}}\Big)^{i+j}\,(-1)^{i-j+1}\,\zeta(2(i+j)-1)\,\frac{\Gamma(2(i+j))}{\Gamma(2i+1)\Gamma(2j+1)}\,,
\end{align}
where $i,j \ge 1$.
They define two semi-infinite matrices whose properties play an important role in what follows.
 
 As mentioned in the Introduction, it proves convenient to interpret the matrix integral \re{Z-Q} as  a partition function of a lattice model defined on 
 the quiver diagram shown in Figure~\ref{fig:qiuv}~\footnote{
Going from \re{Z-Q}, we changed the integration variables as $A_I \to (8\pi^2 N/\lambda)^{-1/2} A_I$. The Jacobian of this transformation coincides with the partition function of $L$ copies of $\mathcal N=4$ SYM. It is not displayed in \re{Z-repr}. This is the reason why \re{Z-repr} gives the difference free energy $Z_{\QQ} =\exp\lr{-\Delta F_{\QQ}}$. \label{foot1}} 
\begin{align}\label{Z-repr}
Z_{\QQ}  = \int \prod_{I=1}^L D A_I\, \exp\lr{-\sum_{I=1}^L\left[ \tr A_I^2    - S_{\rm int}(A_I,A_{I+1})\right]}\,,
\end{align}
where integration goes over $SU(N)$ matrices satisfying   periodic boundary conditions $A_{I+L}=A_I$. The $SU(N)$ matrices $A_I$ (with $I=1,\dots,L$) describe $(N^2-1)$ degrees of freedom living at $L$ sites of the lattice. The second term inside the brackets in the exponent of \re{Z-repr} defines the interaction \re{Sint} between the nearest neighbours on the lattice.  
 
\subsection*{Topological expansion} 

Viewed as a matrix integral, the partition function \re{Z-repr} can be expanded at large $N$ into a sum of  two-dimensional surfaces of different genus. 

A somewhat unusual property of the interaction potential \re{Sint} is that it is given by an infinite sum of double trace terms. Such terms are known to produce touching surfaces \cite{Das:1989fq,Korchemsky:1992tt,Alvarez-Gaume:1992idg,Klebanov:1994pv}. More precisely, the double-trace term of the form $C_{ij} \OO_{i}(A_I) \OO_{j}(A_J)$ generates the touching of two surfaces labelled by $I$ and $J$.
According to \re{Z-repr}, the label $J$ can take three different values $I-1$, $I$ and $I+1$, so that the surface with the label $I=1,\dots, L$ can touch surfaces either of the same type and/or of two adjacent labels (subject to the periodicity condition). 

\begin{figure}[t!]
\begin{center}
\includegraphics[width=1.00\textwidth]{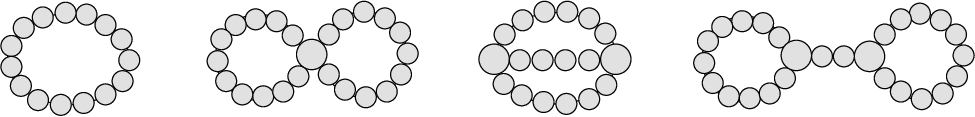}
\caption{Surfaces arising from the topological expansion of the partition function \re{Z-repr} at large $N$. They are obtained by gluing together grey spheres, each carrying the label  $I=1,\dots,L$. The sphere with the label $I$ can touch spheres with the same label and/or with label $I\pm 1$.
}
\label{necklace}
\end{center}
\end{figure}

As a result, the surfaces arising from large $N$ expansion of the partition function \re{Z-repr} 
 take the form of necklaces connected with each other as shown in Figure~\ref{necklace}. The configuration with $h$ holes provides the contribution of order $O(1/N^{2(h-1)})$. For instance, the leftmost diagram in
Figure~\ref{necklace} scales as $O(N^0)$, three remaining diagrams scale as $O(1/N^2)$. Notice that Figure~\ref{necklace} does not contain planar diagrams with $h=0$ (see footnote \ref{foot1}).

In the rest of this section, we develop a technique for computing the contribution of the diagrams shown in Figure~\ref{necklace} to the partition function \re{Z-repr}.

\subsection*{Hubbard-Stratonovich transformation}

Taking an advantage of the symmetry of the partition function \re{Z-repr} under the cyclic shift of matrices, $A_I\to A_{I+1}$, we can simplify 
the interaction term \re{Sint}  by Fourier expanding the single-traces $\OO_{i}(A_I)$ over modes 
with a definite quasimomentum $p_\alpha = {2\pi \alpha/L}$
\begin{align}\label{quasi}
\OO_{i}(A_I) = {1\over \sqrt{L}} \sum_{\alpha=0}^{L-1} e^{ip_\alpha I}\widetilde {\mathcal O}_{\alpha,i} \,,
\end{align}
where $(\widetilde {\mathcal O}_{\alpha,i})^\dagger=\widetilde {\mathcal O}_{-\alpha,i} $ and $i\ge 2$. As we will see in a moment, $p_\alpha$ has the meaning of momenta of excitations propagating across the lattice shown in Figure~\ref{fig:qiuv}.
Inverse relation looks as 
\begin{align}
\widetilde {\mathcal O}_{\alpha,i} = {1\over \sqrt{L}} \sum_{I=1}^L e^{-ip_\alpha I}   \OO_{i}(A_I)\,.
\end{align}
Substituting \re{quasi} into \re{Sint} one gets~\cite{Billo:2021rdb,Billo:2022fnb}
\begin{align}\label{sum-alpha}
\sum_{I=1}^L S_{\rm int}(A_I,A_{I+1}) = \frac12  \sum_{\alpha=0}^{L-1}s_\alpha \,  C_{ij} \widetilde {\mathcal O}_{\alpha,i}\widetilde {\mathcal O}_{-\alpha,j}\,,
\end{align}
 where summation over repeated indices $i,j\ge 2$ is tacitly assumed and the notation was introduced for
 \begin{align}\label{s}
s_\alpha=\sin^2\lr{p_\alpha\over 2}=\sin^2\lr{\pi\alpha\over L}\,.
\end{align}
Notice that $s_\alpha$ vanishes for $\alpha=0$. As a consequence, the corresponding Fourier mode with zero quasimomentum 
$\widetilde {\mathcal O}_{0,i} \sim \sum_{I=1}^L   \OO_{i}(A_I)$
does not contribute to \re{sum-alpha} and, therefore, it is not affected by the interaction.

Following \cite{Beccaria:2023kbl}, we can simplify the matrix integration in \re{Z-repr} by linearizing the double-trace interaction term in \re{sum-alpha}.  This is achieved by introducing auxiliary fields $\widetilde J_{i}(p_\alpha)$ coupled to $\widetilde {\mathcal O}_{-\alpha,i}$ 
\begin{align}\label{int-J}
e^{\sum_{I=1}^L S_{\rm int}(A_I,A_{I+1}) } = \mathcal N \int D\widetilde  J \exp\lr{\sum_{\alpha=1}^{L-1}\left[ \widetilde  J_{i}(p_\alpha) \widetilde {\mathcal O}_{-\alpha,i}-{1\over 2s_\alpha} \,  C_{ij}^{-1} \widetilde J_{i}(p_\alpha)\widetilde J_{j}(-p_\alpha)\right]}\,.
\end{align}
Here the integration measure is $D\widetilde J=\prod_{i,\alpha} d \widetilde J_i(p_\alpha)d \widetilde J_i(-p_\alpha)/\lr{2\pi}$ and the normalization factor is given by $\mathcal N=(\det C)^{-1/2}  (s_1 \dots s_{L-1})^{-1/2}$.

Substituting \re{int-J} into \re{Z-repr} we find that the integral over the matrices $A_I$ factorizes into a product of $L$ independent integrals
\begin{align}\label{prod-Z}
  \int \prod_{I=1}^L D A_I\, \exp\lr{-\sum_{I=1}^L \tr A_I^2  +\sum_{\alpha=1}^{L-1} \widetilde J_{i}(p_\alpha) \widetilde {\mathcal O}_{-\alpha,i}  }
= Z(J_1) \dots Z(J_L)\,,
\end{align}
where $Z(J_I)$ is a partition function of a Gaussian $SU(N)$ matrix model  
\begin{align}\notag\label{gf}
{}& Z(J_I)  = \int DA \, \exp\lr{-\tr A^2 + J_{I,i}\, \OO_{i}(A)}
\\
{}& \phantom{Z(J)}=\exp\lr{N J_{I,i} G_i + \frac1{2!} J_{I,i_1}  J_{I,i_2}  G_{i_1i_2} + \frac1{3! N } J_{I,i_1}J_{I,i_2}J_{I,i_3}G_{i_1i_2i_3}+\dots }\,.
\end{align}
and the sources $J_{i,I}$ (with $i\ge 2$) are given by Fourier series analogous to \re{quasi}
\begin{align}\label{sour}
J_{I,i} = {1\over \sqrt{L}} \sum_{\alpha=1}^{L-1} e^{ip_\alpha I}\widetilde J_{i}(p_\alpha)\,.
\end{align}
Notice that the sum in this relation does not contain the term with $\alpha=0$ and, as a consequence, the sources satisfy the relation $\sum_{I=1}^L J_{i,I}=0$.

The exponent in \re{gf} is given by a linear combination of (connected part of) correlation functions in a Gaussian matrix model 
\begin{align}\label{G} 
G_{i_1 \dots i_n} {}&= N^{n-2} \vev{\OO_{i_1}(A) \dots \OO_{i_n}(A)}_{0,c}
 = G_{i_1 \dots i_n} ^{(0)} + {1\over N^2} G_{i_1 \dots i_n} ^{(1)} +\dots \,.
\end{align}
where the subscript `$0$' in the first relation indicates that the expectation value is evaluated with a Gaussian measure.
Here the factor of $N^{n-2}$ was inserted for convenience, it ensures that $G_{i_1 \dots i_n}$ stays finite for $N\to\infty$. 
The second relation in \re{G} yields the expansion of the correlator at large $N$.  We do not present here the explicit expressions for 
$G_{i_1 \dots i_n} ^{(0)}$ and $G_{i_1 \dots i_n} ^{(1)}$, they can be found in Appendix~A of \cite{Beccaria:2023kbl}.

Combining together the relations \re{int-J}, \re{gf} and \re{prod-Z}, we arrive at the following representation of the partition function  \re{Z-repr}
\begin{align}\label{Z-S}
Z_{\QQ}  =  \mathcal N \int D\widetilde  J\exp\lr{ -\sum_{\alpha=1}^{L-1}{1\over 2s_\alpha} \,  C_{ij}^{-1} \widetilde J_{i}(p_\alpha)\widetilde J_{j}(-p_\alpha)+ \sum_{n=1}^\infty {1\over N^{n-2} n!}S_n(J)}\,,
\end{align}
where the second term in the exponent is given by a linear combination of  homogenous polynomials in $J$'s with the coefficients defined by the correlators \re{G}
\begin{align}\label{Sn} 
S_n(J) {}&= G_{i_1\dots i_n} \sum_{I=1}^L J_{I,i_1} \dots J_{I,i_n} \,.
\end{align}
The exponent of \re{Z-S} has large $N$ expansion analogous to that of the correlator \re{G}. 

The function $S_n(J)$ has  
the meaning of an effective action for the sources $J_I$ induced by the double trace interaction \re{Sint}.  At large $N$ the leading contribution to the exponent of \re{Z-S} comes from $S_1(J)$. By definition \re{Sn}, it is proportional to the sum of sources at all sites and, therefore, it vanishes
$S_1(J)= G_i\sum_{I=1}^L J_{I,i}=0$ (see \re{sour}).
We can apply \re{sour} to express the remaining $S_n(J)$ with $n\ge 1$ in terms of the sources $\widetilde J_{i}(p_\alpha)$
\begin{align}\notag\label{S's}
S_2 {}&= \sum_{\alpha=1}^{L-1}G_{i_1i_2}\widetilde J_{i_1}(p_\alpha)\widetilde J_{i_2}(-p_\alpha)\,,
\\
S_n {}&={1\over L^{n/2-1}} \sum_{\alpha_1,\dots,\alpha_n=1}^{L-1} G_{i_1 \dots i_n}
\widetilde J_{i_1}(p_{\alpha_1})\dots \widetilde J_{i_n}(p_{\alpha_n})\delta\lr{p_{\alpha_1}+\dots +p_{\alpha_n}}\,,
\end{align}
where the $\delta-$function imposes the condition of conservation of the quasimomentum
\begin{align}\label{delta}
p_{\alpha_1}+\ldots + p_{\alpha_n} = 0 \quad \text{mod$(2\pi)$}\,.
\end{align}
To simplify formulae we will use a short-hand notation for the sum over quasimomenta in \re{S's}
\begin{align}
S_n \equiv {1\over L^{n/2-1}}  G_{i_1 \dots i_n} \widetilde J_{i_1} \star \dots  \star  \widetilde J_{i_n}\,.
\end{align}

Finally, we substitute \re{S's} into \re{Z-S} to obtain the following representation of the partition function
\begin{align}\label{Z-Feynman}
Z_{\QQ}  =  \mathcal N \int D\widetilde  J\exp\lr{ -\frac12 \sum_{\alpha=1}^{L-1}\lr{{1\over s_\alpha} \,  C_{ij}^{-1} -G_{ij}}\widetilde J_{i}(p_\alpha)\widetilde J_{j}(-p_\alpha)+ \sum_{n=3}^\infty {G_{i_1 \dots i_n} \over n! (N^2 L)^{n/2-1}}
 \widetilde J_{i_1} \star \dots  \star  \widetilde J_{i_n}}\,.
\end{align} 
This relation depends on two sets of semi-infinite matrices -- the expansion coefficients \re{C} and the correlation functions \re{G} in a Gaussian $SU(N)$ matrix model.

The first sum in the exponent of \re{Z-Feynman} is quadratic in the sources $ \widetilde J_{i}(p_\alpha)$. It describes a propagation of excitations with the quasimomentum $p_\alpha$ along the quiver diagram in Figure~\ref{fig:qiuv}. The second sum in \re{Z-Feynman} describes the interaction between these excitations. The coupling of $n$ excitations is proportional to the correlation functions $G_{i_1 \dots i_n}$ and it is suppressed by the factor of $1/N^{n-2}$. As a consequence, in the leading large $N$ limit  the partition function \re{Z-Feynman} is given by a Gaussian integral  
\begin{align}\label{lo-det}
Z_{\QQ} =  {1 \over \sqrt {\det\left( 1-  s_\alpha \,  C  G \right)}} + O(1/N^2) \,, 
\end{align}
where the normalization factor $\mathcal N$ was replaced with its expression \re{int-J}.
Subleading corrections in $1/N$ can be obtained by expanding \re{Z-Feynman} in powers of the interaction term. 
 
\subsection*{Feynman diagram technique}

The partition function \re{Z-Feynman} can be evaluated using a Feynman diagram technique. The first term in the exponent of \re{Z-Feynman} defines a propagator of the fields $\tilde J_i(p_\alpha)$
\begin{align}\label{prop}
\vev{\widetilde J_{i}(p_{\alpha}) \widetilde J_{j}(p_{\alpha'}} =
X_{ij}(p_\alpha)\delta(p_\alpha + p_{\alpha'})\,,
\end{align}
where the $\delta-$function imposes the condition \re{delta} and semi-infinite matrix $X_{ij}$ (with $i,j\ge 2$) is defined as
\begin{align}\label{X}
X_{ij}(p_\alpha) =  \left[ s_\alpha  C  \lr{1 -s_\alpha G C}^{-1}\right]_{ij}\,.
\end{align}
We recall that $p_\alpha=2\pi \alpha/L$ is the quasimomentum of excitations (with $\alpha=0,\dots,L-1$). Taking into account \re{s} we note that $X_{ij}(p_\alpha)$ vanishes for $\alpha=0$ and, therefore, the field $\widetilde J_{i}(p_{\alpha})$ with $p_\alpha=0$ does not propagate. 

\begin{figure}[t!]
\psfrag{p}[cc][cc]{$p_\alpha$}\psfrag{p1}[cc][cc]{$p_{\alpha_1}$}\psfrag{p2}[cc][cc]{$p_{\alpha_2}$}\psfrag{p3}[cc][cc]{$p_{\alpha_3}$}
\psfrag{0}[cc][cc]{$0$}
\psfrag{a}[cc][cc]{(a)}\psfrag{b}[cc][cc]{(b)}\psfrag{c}[cc][cc]{(c)}\psfrag{d}[cc][cc]{(d)}
\begin{center}
\includegraphics[width=1.00\textwidth]{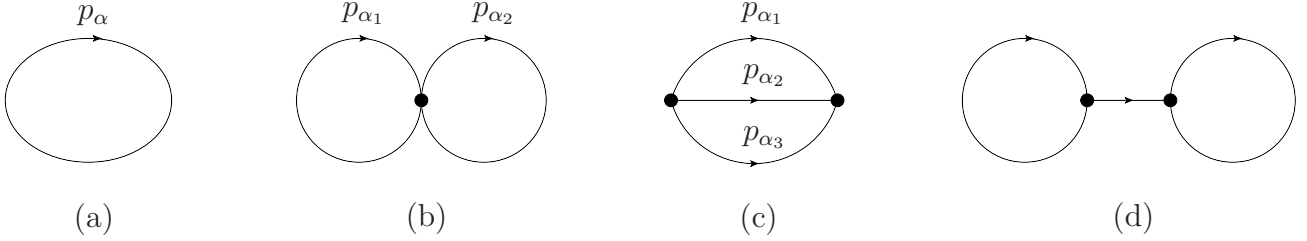}
\caption{The diagrams contributing to the free energy up to order $O(1/N^2)$. Solid lines represent the propagator \re{prop}, black dots denote the interaction vertices \re{S's}.
The diagram (d) contains a line with zero momentum exchange and produces a vanishing contribution.
}
\label{fig}
\end{center}
\end{figure}

The free energy $\Delta F_{\QQ}=-\log Z_{\QQ}$ is given by the sum of `vacuum' Feynman diagrams shown in Figure~\ref{fig}. They involve the propagators \re{prop} and the interaction vertices generated by the second term in the exponent of \re{Z-Feynman}. The contribution of a diagram with $h$ loops scales as $O(1/N^{2(h-1)})$. The diagrams in Figure~\ref{fig} are in the one-to-one correspondence with those shown in Figure~\ref{necklace}. The former diagrams can be obtained from the latter by simply replacing the chains of touching surfaces by solid lines.

The diagram in Figure~\ref{fig}(d) contains an exchange with zero quasimomentum.  It produces a vanishing contribution to $\Delta F_{\QQ} $. The contribution of the three remaining diagrams reads
\begin{align}\notag\label{F-free}
\Delta F_{\QQ}  {}&= \frac12\sum_{\alpha_1=1}^{L-1}\log\det(1- s_{\alpha_1} C G)
\\\notag
{}& -{1\over 8N^2 L}\sum_{\alpha_1, \alpha_2=1}^{L-1}   G_{i_1i_2i_3i_4}X_{i_1i_2}(p_{\alpha_1}) X_{i_3i_4}(p_{\alpha_2})  
\\
{}& -{1\over 12N^2 L}\sum_{\alpha_1, \alpha_2=1}^{L-1}  G_{i_1i_2i_3}G_{j_1j_2j_3} X_{i_1j_1}(p_{\alpha_1}) X_{i_2j_2}(p_{\alpha_2})
X_{i_3j_3}(-p_{\alpha_1}-p_{\alpha_2}) \,,
\end{align}
where summation over repeated indices is tacitly assumed.
This relation is valid up to  $O(1/N^4)$  corrections.
The first term in \re{F-free} comes from the diagram shown in Figure~\ref{fig}(a). Its contribution to $\Delta F_{\QQ}$ can be read from \re{lo-det}. The second and third terms in \re{F-free} come from the diagrams shown in Figure~\ref{fig}(b) and \ref{fig}(c), respectively. They are given by the product of the propagators \re{X} and the correlation functions \re{G} summed over the quasimomenta propagating inside the loops.

The relation \re{F-free} holds for arbitrary $L$. As explained in the Introduction, $\Delta F_{\QQ}$ has to vanish for $L=1$ to any order in $1/N$. This property holds separately for each term in \re{F-free}. 
For $L=2$ the sums in \re{F-free} contain only one term with $\alpha_1=\alpha_2=1$.
In addition, the last term in \re{F-free} vanishes because the propagator $X_{i_3j_3}(-p_{\alpha_1}-p_{\alpha_2})$ carries vanishing quasimomentum $-p_{\alpha_1}-p_{\alpha_2}=0$ (mod  $2\pi$). 
The two remaining terms in \re{F-free} as well as higher order corrections in $1/N$ were computed in \cite{Beccaria:2023kbl}. In the next section, we evaluate \re{F-free} for arbitrary $L$.

\section{Large $N$ expansion of the free energy}\label{sect3}

The relations \re{X} and \re{F-free} involve two sets of semi-infinite matrices, $C_{ij}$ and $G_{i_1\dots i_n}$, defined in \re{C} and \re{G}, respectively. The matrix $C_{ij}$   depends on 't Hooft coupling $\lambda$ 
and is independent on $N$. 
The correlators $G_{i_1\dots i_n}$ are independent on $\lambda$ and admit the expansion \re{G} in powers of $1/N^2$.

Replacing the correlators $G_{i_1\dots i_n}$ in \re{F-free} with their large $N$ expansion \re{G}, we match \re{F-free} to \re{F-exp} and identify the first two terms of the expansion of the free energy,  
\begin{align}\label{F0}
F_L^{(0)} ={}& \frac12\sum_{\alpha=1}^{L-1}\log\det(1- s_\alpha C G^{(0)} )\,,
\\\notag\label{F1}
F_L^{(1)} ={}& -\frac12\sum_{\alpha=1}^{L-1} G_{i_1i_2}^{(1)} X^{(0)} _{i_1i_2}(p_\alpha) 
 -{1\over 8 L}\sum_{\alpha_1, \alpha_2=1}^{L-1}   G^{(0)} _{i_1i_2i_3i_4}X^{(0)} _{i_1i_2}(p_{\alpha_1}) X^{(0)} _{i_3i_4}(p_{\alpha_2})  
\\
{}& -{1\over 12 L}\sum_{\alpha_1, \alpha_2=1}^{L-1}  G^{(0)} _{i_1i_2i_3}G^{(0)} _{j_1j_2j_3} X^{(0)} _{i_1j_1}(p_{\alpha_1}) X^{(0)} _{i_2j_2}(p_{\alpha_2})
X^{(0)} _{i_3j_3}(-p_{\alpha_1}-p_{\alpha_2}) \,.
\end{align}
Here $G_{i_1i_2\dots}^{(0)}$ and $G_{i_1i_2\dots}^{(1)}$ are the first two terms of the large $N$ expansion of the correlator \re{G}. The term proportional to $G_{i_1i_2}^{(1)}$  arises in \re{F1} from the expansion of the first term in \re{F-free} to order $O(1/N^2)$. The propagator $X^{(0)} _{i_1i_2}(p_\alpha)$ is given by \re{X} with the two-point correlator $G_{i_1i_2}$ replaced by its leading large $N$ expression  $G^{(0)}_{i_1i_2}$.

\subsection{Leading nonplanar correction}
 
We recall that the matrix elements $C_{ij}$ vanish for indices $i$ and $j$ of different parity. 
Being the two-point correlator in a Gaussian matrix model, the semi-infinite matrix $G_{ij}$
has the same property. 
In a close analogy with \re{C} we can define nonzero matrix elements $\mathsf Q^+_{ij}\equiv G^{(0)}_{2i,2j}$ and $\mathsf Q^-_{ij}\equiv G^{(0)}_{2i+1,2j+1}$ (with $i,j\ge 1$). Then, the relation \re{F0} can be simplified as
\begin{align}\label{F0-ini}
F_L^{(0)}=\frac12\sum_{\alpha=1}^{L-1}
\Big[ \log\det(1- s_\alpha \mathsf Q^+ C^+ )+\log\det(1- s_\alpha \mathsf Q^- C^-)\Big] \,,
\end{align}
where the semi-infinite matrices $C^\pm$ were defined in \re{C}.
Notice that the dependence of $F_L^{(0)}$ on the number of nodes $L$ enters through $s_\alpha$ defined in \re{s}. 

The properties of semi-infinite matrices $\mathsf Q^+ C^+$ and $\mathsf Q^- C^-$ were discussed in \cite{Beccaria:2023kbl}. 
Both matrices are related by a similarity transformation to the same, universal matrix $K_{\ell}(\chi)$ evaluated for $\ell=1,2$ 
\begin{align}\notag\label{simil}
{}& \mathsf Q^+ C^+ = U^+ K_{\ell=1}(\chi) (U^+)^{-1}\,, 
\\[2mm]
{}& \mathsf Q^- C^- = U^- K_{\ell=2}(\chi)(U^-)^{-1}\,,
\end{align}
Its matrix elements are given by integral of the product of two Bessel functions of the first kind
\begin{align}\label{K-mat}
\left(K_\ell(\chi)\right)_{ij} =2(-1)^{i+j} \sqrt{2i+\ell-1}\sqrt{2j+\ell-1}\int_0^\infty {dx\over x} J_{2i+\ell-1}(x)J_{2j+\ell-1}(x)
\chi\lr{2\pi x\over \sqrt\lambda}\,,
\end{align}
where $i,j\ge 1$. The  function $\chi(x)$ is conventionally called the symbol of the matrix. For the matrices \re{simil} it is given by
\begin{align}\label{chi}
\chi(x) = -{1\over\sinh^2(x/2)}\,.
\end{align}

The expression on the right-hand side of \re{F0-ini} can be expressed in terms of a determinant of the semi-infinite matrix \re{K-mat}
\begin{align}\label{F-ell}
\mathcal F_\ell(\chi) = \log\det(1-K_\ell(\chi))\,.
\end{align}
Indeed, the matrices $(1-s_\alpha \mathsf Q^\pm C^\pm)$ entering \re{F0-ini} are related by the similarity transformation \re{simil} to the matrices $(1-s_\alpha  K_{\ell=1,2}(\chi))$ and, as a consequence, their determinants coincide. Taking into account that $s_\alpha  K_{\ell}(\chi)=K_{\ell}(s_\alpha  \chi)$  
we obtain from \re{F0-ini}  
\begin{align}\label{F0-sum}
F_L^{(0)}=\frac12\sum_{\alpha=1}^{L-1}\Big[\mathcal F_{\ell=1}(s_\alpha \chi)+\mathcal F_{\ell=2}(s_\alpha \chi)\Big]\,,
\end{align}
where the function $\mathcal F_{\ell}(s_\alpha \chi)$ is given by \re{F-ell} with the symbol $\chi(x)$ replaced with $s_\alpha \chi(x)$.
 
For $L=2$ the relation \re{F0-sum} agrees with analogous relation obtained in \cite{Beccaria:2023kbl}.

\subsection{Next-to-leading nonplanar correction}

The next-to-leading nonplanar $O(1/N^2)$ correction to the free energy is given by  \re{F1}. Various terms in
\re{F1} involve the product of semi-infinite matrices defined in \re{X} and  \re{G}.  
They can be efficiently computed using the technique developed in \cite{Beccaria:2023kbl}. 

It proves convenient to introduce the so-called Bessel operator
\begin{align}\label{K-op}
{\bm K_\ell}=\sum_{i,j\ge 1} \ket{\psi_i}\left(K_\ell(\chi)\right)_{ij}\bra{\psi_j}\,,
\end{align}
where $\left(K_\ell(\chi)\right)_{ij}$ is given by \re{K-mat} and the functions $\psi_i(x)$ (with $i\ge 1$) form an orthonormal basis 
\begin{align}\notag\label{psi}
& \psi_i(x) = (-1)^i \sqrt{2i+\ell-1}{J_{2i+\ell-1}(\sqrt x)\over \sqrt x}\,,
\\
& \vev{\psi_i|\psi_j} = \int_0^\infty dx\, \psi_i(x) \psi_j(x) = \delta_{ij}\,.
\end{align}
Then, the product of semi-infinite matrix \re{K-mat} can be evaluated by taking matrix elements of a power of the Bessel operator, e.g. $\left(K^n_\ell(\chi) \right)_{ij} = \vev{\psi_i|{\bm K_\ell}^n|\psi_j}$. This property allows us to express \re{F-ell} in terms of a Fredholm determinant of the Bessel operator
\begin{align}\label{Fred}
\mathcal F_\ell(\chi) =\log\det(1-\bm K_\ell)\,.
\end{align}
In the similar manner, the infinite sums in \re{F1} can be expressed in a concise form in terms of matrix elements of the Bessel operator \re{K-op}.

For $L=2$ the infinite sums in \re{F1} were computed in \cite{Beccaria:2023kbl}. In this case they can be expressed in terms of matrix elements of the resolvent of the Bessel operator   
\begin{align}\label{w-nm}
w_{nm}(\chi)= \vev{\phi_n|\bm{\chi} {1\over 1-\bm K_\ell} |\phi_m}\,,
\end{align}
where the matrix element is evaluated over the states
$\phi_n(x)$ (with $n\ge 0$) defined as
\begin{align}\label{phi}
\phi_0(x) = J_\ell(\sqrt x)\,,\qqqquad 
\phi_n(x)= (x\partial_x)^n \phi_0(x)\,.
\end{align}
The operator $\bm{\chi}$ in \re{w-nm} has a diagonal kernel and it acts on a test function as $\bm{\chi} f(x) = \chi(2\pi \sqrt {x/\lambda})f(x)$.  

For arbitrary $L\ge 3$, the evaluation of \re{F1} goes along the same lines as in \cite{Beccaria:2023kbl}. 
In a close analogy with \re{F0-sum}, we use \re{w-nm} to define the $L-$dependent matrix elements
\begin{align}\label{w}
w_{nm}(s_\alpha \chi)= s_\alpha \vev{\phi_n|\bm{\chi} {1\over 1-s_\alpha  \bm K_\ell} |\phi_m}\,,
\end{align}
where $s_\alpha$ is given by \re{s}. Because the operators $\bm{\chi}$ and $\bm K_\ell$ are linear in $\chi$, the additional factor of $s_\alpha$ in \re{w} can be absorbed into redefinition of the symbol  
\begin{align}\label{chi-sub}
\chi(x) \to s_\alpha \chi(x)\,.
\end{align}
Applying this transformation to \re{w-nm}, we arrive at \re{w}.

Following \cite{Beccaria:2023kbl}, we replace the correlation functions $G^{(0)}$ and $G^{(1)}$ in \re{F1} with their explicit expressions and 
express infinite sums in \re{F1} in terms of matrix elements \re{w} evaluated for $\ell=1$ and $\ell=2$.~\footnote{We refer the interested reader to  \cite{Beccaria:2023kbl} for details of the calculation.}  Roughly speaking, each propagator 
$X^{(0)}_{ij}(p_\alpha)$ in \re{F1} gives rise to a linear combination of matrix elements $w_{nm}(s_\alpha\chi)$ with different $n$ and $m$. 
Then, the three sums on the right-hand side of \re{F1} become homogenous polynomials  in matrix elements \re{w} of degree $1$, $2$ and $3$, respectively. Going through the calculation we get
\begin{align}\notag\label{F1-QL}
F_L^{(1)}{}&= \frac{\vvev{w^+_{0,0}}}{384} 
+\frac{13
   \vvev{w^+_{0,1}}}{96}
   -\frac{\vvev{w^+_{0,2}}}{48}-\frac{\vvev{w^+_{1,1}}}{96}
+\frac{
   13\vvev{w^-_{0,0}}}{96} +\frac{5
  \vvev{w^-_{0,1}}}{48} -\frac{\vvev{w^-_{0,2}}}{48} -\frac{\vvev{w^-_{1,1}}}{96}
\\[1.5mm] \notag
{}&- {2\over L} \Big[\frac{1}{64} \vvev{w^+_{0,0}+w^-_{0,0}}\vvev{w^+_{0,1}+w^-_{0,1}}+\frac{1}{128} \vvev{w^+_{0,0}} \vvev{w^-_{0,0}}
+\frac{1}{64}\vvev{w^-_{0,0}}^2+\frac{1}{256} \vvev{w^+_{0,0}}^2\Big]
\\[1.5mm]
&  
-{1\over 768 L}\sum_{\alpha_1, \alpha_2=1}^{L-1}  
\Big(w^+_{0,0}(s_{\alpha_1}\chi)w^+_{0,0}(s_{\alpha_2}\chi)
 + 3w^-_{0,0}(s_{\alpha_1}\chi)w^-_{0,0}(s_{\alpha_2}\chi)\Big)w^+_{0,0}(s_{-\alpha_1-\alpha_2}\chi)\,,
\end{align}
where the notation was introduced for
\begin{align}\label{w-vev}
\vvev{w^+_{nm}} = \sum_{\alpha=1}^{L-1} w_{nm}(s_\alpha\chi)\Big|_{\ell=1}\,,\qqqquad
\vvev{w^-_{nm}} = \sum_{\alpha=1}^{L-1} w_{nm}(s_\alpha\chi)\Big|_{\ell=2}\,.
\end{align}
The first sum on the right-hand side of \re{F1} is given by expression on the first line of \re{F1-QL}. In a similar manner, the two remaining sums in \re{F1} give rise to the second and third lines of \re{F1-QL}.  

For $L=2$ the relation \re{F1-QL} coincides with the analogous expression in the $\mathsf Q_2$ model derived in \cite{Beccaria:2023kbl} (see Eq.~(4.16) there). In this case,  the last line of \re{F1-QL} vanishes because $s_{-\alpha_1-\alpha_2}=0$ for $\alpha_1=\alpha_2=1$.  

\subsection{Nonplanar corrections at weak and strong coupling}

The relations \re{F0-sum} and \re{F1-QL} provide the expressions for the nonplanar corrections to the free energy in terms of the Fredholm determinant of the Bessel operator  and the matrix elements of its resolvent, defined in \re{F-ell} and \re{w-nm}, respectively. We would like to emphasize that these relations hold for an arbitrary 't Hooft coupling. 

\subsection*{Weak coupling}

At weak coupling, the calculation of \re{F0-sum} and \re{F1-QL} simplifies significantly by noticing that the matrix elements \re{K-mat} vanish for $\lambda\to 0$. To see this, 
one changes the integration variable in \re{K-mat} as $x\to g x$ and replaces the Bessel functions with their small $g$ expansion  
to find that $\left(K_\ell(\chi)\right)_{ij} =O(\lambda^{\ell+i+j-1})$. 
This property allows us to expand \re{F-ell} in powers of $K_\ell$ 
\begin{align}\label{trs}
\mathcal F_\ell(\chi) = -\tr K_\ell(\chi) -\frac12 \tr K^2_\ell(\chi)  -\frac13 \tr K^3_\ell(\chi) + O(\lambda^{4(\ell+1)})
\end{align}
and, then, replace the semi-infinite matrix \re{K-mat} with its finite-dimensional minor. For the matrix elements \re{w-nm}, the analogous expansion is worked out in Appendix~\ref{app} (see \re{m.e.}). 

Substituting the resulting expressions of $\mathcal F_\ell(\chi)$ and $w_{nm}(\chi)$  into \re{F0-sum} and \re{F1-QL}, we find after some algebra
\begin{align}
\label{F-weak}\notag
F_L^{(0)}&{}= \frac{3\zeta(3)}{64\pi^{4}}\sigma_{2}(L)\,\l^{2}-\frac{15\zeta(5)}{512\pi^{6}}\sigma_{2}(L)\,\l^{3}
+\bigg(\frac{315\zeta(7)}{16384\pi^{8}}\sigma_{2}(L)-\frac{9\zeta(3)^{2}}{4096\pi^{8}}\sigma_{4}(L)\bigg)\,\l^{4}
\\
&{} +\bigg( -\frac{441\zeta(9)}{32768\pi^{10}}\sigma_{2}(L)+\frac{15\zeta(3)\zeta(5)}{4096\pi^{10}}\sigma_{4}(L)\bigg)\,\l^{5}
+O(\lambda^6)\,,
\\[2mm]
\notag\label{F1-weak}
F_L^{(1)}{}&= -\frac{3   \zeta (3)}{64 \pi ^4}\sigma_2(L)\lambda ^2+\frac{25  \zeta (5)}{512 \pi
   ^6}\sigma_2(L)\lambda ^3  
\\
{}&   
   +  \left( -\frac{735 \zeta (7)}{16384 \pi
   ^8}\sigma_2(L)+\frac{9 \zeta (3)^2}{2048 \pi ^8}\left(\sigma_4(L)-{3\over L} \sigma_2^2(L)\right)\right) \lambda ^4 + O(\lambda^5)\,,
\end{align}
where the dependence on the number of nodes $L$ enters through the functions
\be\label{sigma}
\sigma_{2n}(L) =\sum_{\alpha=1}^{L-1} s_\alpha^n= \sum_{\alpha=1}^{L-1}\sin^{2n}\left(\frac{\pi \alpha}{L}\right)\,.
\ee
Being combined together, the relations \re{F-weak} and \re{F1-weak} yield the weak coupling expansion of the free energy \re{F-exp}.  

The following comments are in order.

The terms proportional to $\sigma_{2n}(L)$ arise in \re{F-weak} and \re{F1-weak} from the expansion of the determinants and matrix elements in powers of $K_{\ell}(s_\alpha\chi)=s_\alpha  K_{\ell}(\chi)$. The odd Riemann zeta values come from 
the integrals $\int_0^\infty dx\, x^{2n+1} \chi(x)=-4 (2n+1)! \zeta(2n+1)$ that arise in the small $\lambda$ expansion of  \re{K-mat}.

To high orders in the coupling constant, the leading function $F_L^{(0)}$ is given by multi-linear combinations of odd zeta values multiplied by $\sigma_{2n}(L)$ (see \re{weak-gen} below). As compared with the leading large $N$ contribution \re{F-weak}, the function $F_L^{(1)}$ contains  terms like $\sigma_2^2(L)/L$ that are bilinear in
$\sigma$. They come from the diagrams shown in Figure~\ref{fig}(b) and (c).  

Let us examine the dependence of \re{sigma} on $L$. For lowest values of $n$, 
it is straightforward to check that $\sigma_2(L)=L/2$ for $L\ge 2$, $\sigma_4(L)$ equals $1$ for $L=2$ and $3L/8$ for $L\ge 3$, and so on. 
The general expression of $\sigma_{2n}(L)$
reads  \cite{fonseca2017basic} 
\be\label{sigma-L}
\sigma_{2n}(L) = \begin{cases} \displaystyle
\  2^{1-2n}\,L\,\bigg[\binom{2n-1}{n-1}+\sum_{p=1}^{\lfloor n/L\rfloor}(-1)^{pL}\binom{2n}{n-pL}\bigg], & n\ge L, 
\\[4mm]
 \displaystyle
\ 2^{1-2n}\,L\,\binom{2n-1}{n-1}, & n<L.
\end{cases}
\ee
Notice that $\sigma_{2n}(L)$ is a linear function of $L$ for $n<L$. This property plays an important role in the next section where we discuss the asymptotic behaviour of the free energy at large $L$.  

For arbitrary $n$, the function $\sigma_{2n}(L)$ has different $L-$dependence for $L\le n$ and  $L>n$. The change of the behaviour occurs at $n=L$ and the corresponding function $\sigma_{2L}(L)$ first appears in the weak coupling expansion of the free energy \re{F-weak} at order $O(\lambda^{2L})$. We show in section \ref{sect:large} that this fact has an interesting interpretation in terms of wrapping corrections in the quiver lattice model.     
 
\subsection*{Strong coupling}

To derive the strong coupling expansion of the free energy \re{F0-sum} and \re{F1-QL}, we make use of the known properties of the Bessel operator \re{K-op}. 

We recall that the leading nonplanar correction to the free energy \re{F0-sum} involves the Fredholm determinant of this operator \re{Fred}. 
 The strong coupling expansion of $\mathcal F_\ell(\chi)$ for generic symbol $\chi(x)$  was derived in \cite{Belitsky:2020qrm,Belitsky:2020qir}. The first few terms of the expansion are given by 
 \begin{align} \notag\label{F-ell-exp}
 \mathcal F_\ell(\chi) ={}& -2g I_0(\chi)  - \frac12(2\ell-1) \log g +B_\ell(\chi) 
\\\notag
{}&
-\frac1{16g} (2\ell-1)(2\ell-3) I_1(\chi)  -\frac1{64g^2} (2\ell-1)(2\ell-3) I_1^2(\chi)  
\\
{}& 
-\frac{1}{3072 g^3}(2 \ell-1) (2 \ell-3)\left[(2 \ell-5) (2 \ell+1)I_2(\chi) +16 I_1^3(\chi) \right] + O(1/g^4)
 \,,
\end{align}
 where the notation was introduced for 
\begin{align}\label{g}
g={\sqrt\lambda\over 4\pi}\,.
\end{align}
The constant term $B_\ell(\chi)$ is conventionally called the Widom-Dyson constant. Its explicit expression can be found in Appendix~\ref{app} (see \re{B-const}). The functions $I_n(\chi)$ (with $n\ge 0$) are defined as
\begin{align}\label{In}
I_n(\chi) = {1\over (2n-1)!!} \int_0^\infty {dx\over\pi} (x^{-1}\partial_x)^n x\partial_x\log\left(1-\chi(x)\right) \,.
\end{align}
High order corrections to \re{F-ell-exp} are given by multi-linear combinations of $I_n(\chi)$.  

Substituting \re{F-ell-exp} into \re{F0-sum} we encounter the functions $I_n(s_\alpha \chi)$.
Replacing $\chi(x)$ and $s_\alpha$ in \re{In} with their expressions, \re{chi} and \re{s}, respectively, and going through the calculation 
we find
\begin{align}\notag\label{I0}
& I_0(s_\alpha \chi)=-2\pi {\alpha(L-\alpha)\over L^2}\,,
\\[2mm] 
& I_1(s_\alpha \chi) =-\frac1{2 \pi }\left[{\psi\left(\frac{\alpha }{L}\right)+\psi\left(1-\frac{\alpha
   }{L}\right)-2 \psi(1) }\right]\,,
\end{align}
where $\psi(x) = d  \log\Gamma(x)/dx$ is the Euler $\psi-$function. For arbitrary $n\ge 1$ we have  
\begin{align}\label{In1}
I_n(s_\alpha \chi)  
   = {(-1)^n  \over (2\pi)^{2n-1}(2n-2)!} \left[ \psi ^{(2n-2)}\left(\frac{\alpha }{L}\right)+\psi
   ^{(2n-2)}\left(1-\frac{\alpha }{L}\right)-2\psi ^{(2n-2)}(1)\right],
\end{align}
 where $ \psi^{(n)}(x)= (d/dx)^n \psi(x)$. 
 
To find the nonplanar correction to the free energy \re{F1-QL}, we use the strong coupling expansion of the matrix elements \re{w-nm} derived in 
\cite{Beccaria:2023kbl}. The matrix element $w_{00}(\chi)$  is given by a derivative of the function  \re{F-ell} with respect to the coupling constant
\begin{align}\label{w00}
w_{00}(\chi) =  - 4\lambda  \partial_\lambda \mathcal F_\ell(\chi)\,.
\end{align}
The remaining matrix elements can be found by taking into account functional relations  
\begin{align}\label{fun-rel}w_{nm} =w_{mn}\,,\qquad  
\lr{\frac12 g\partial_g-1}w_{nm} =\frac14 w_{0n}w_{0m} + w_{n+1,m} + w_{n,m+1}\,.    
\end{align}
Note that the relations \re{w00} and \re{fun-rel} hold for arbitrary 't Hooft coupling $\lambda$. 
The relations \re{fun-rel} allow us to express $w_{nm}(\chi)$ for any $n$ and $m$ in terms of independent quantities $w_{0,2n}$ (with $n=0,1,\dots$). The  strong coupling expansion of $w_{00}, w_{01}, w_{11}$ and $w_{02}$ is given in Appendix~\ref{app}, see \re{w-expr}.

Substituting the obtained expressions for $ \mathcal F_\ell(s_\alpha \chi) $ and $w_{nm}(s_\alpha\chi)$ into \re{F0-sum} and \re{F1-QL}, we find after some algebra the following expression for the nonplanar corrections to the free energy
\begin{align}\label{FL-free}
F_L^{(0)}={}& \frac{2\left(L^2-1\right)}{3 L}g\pi - (L-1) \log g + \frac12 C_L^{(0)} -{ L \log L \over 16 \pi g}  + O(1/g^2)\,,
\\[2mm]
\notag\label{F1-res}
F_L^{(1)} ={}& -(\pi g)^3 \frac{(L^2-1) (L^2+1)}{30 L^3}+(\pi g)^2 \frac{(L^2-1) (L^2-9)}{360 L^3}
\\
{}&
+ (\pi  g)\frac{5 (L^2-1)}{384 L}+ (\pi  g)\frac{C_L^{(1)}}{48} 
+O(g^0) \,,
\end{align}
where $g$ is defined in \re{g}. The details of the calculation can be found in Appendix~\ref{app}. 

The coefficient function $C_L^{(0)}$  in \re{FL-free} is given by a linear combination of the Widon-Dyson constant $B_\ell(s_\alpha\chi)$
entering \re{F-ell-exp}. Its explicit expression can be found in Appendix~\ref{app}, see \re{app:C}.  
For lowest values of $L$ it looks as
\begin{align}\notag\label{C-vals}
& C^{(0)}_{L=2}=-12 \log \mathsf A + 1-\frac83 \log 2  \,,
\\[2mm] 
&  C^{(0)}_{L=3}=-16\log \mathsf A+\frac{4}{3}-\frac{17}{6}\log (3)+2 \log {\Gamma \left(\ft{2}{3}\right)\over \Gamma
   \left(\ft{1}{3}\right)} \,,
\end{align}  
where $\mathsf A$ is the Glaisher constant.  
At large $L$ we find that $C_L$ grows linearly with $L$
\begin{align}\label{C-large}
C^{(0)}_L=L\lr{\frac12-12\log\mathsf A} - 2\log L + \log (4\pi)  +O(1/L)\,.
\end{align}
The coefficient function $C ^{(1)}_L$ in \re{F1-res} is given by
\begin{align}
C ^{(1)}_L=\sum_{\alpha=1}^{L-1}  { {\alpha\over L} \left(1 -{\alpha\over L}\right)} \left[ \psi  \left(\frac{\alpha }{L}\right)+\psi  \left(1-\frac{\alpha
   }{L}\right)-2 \psi  (1)  \right].
\end{align}
At large $L$ it behaves as
\begin{align}
C ^{(1)}_L = -L \lr{4 \log \mathsf A-\frac{\gamma }{3}} +1+ O(1/L)\,.
\end{align}
where $\gamma$ is the Euler constant.
 
The relations \re{FL-free} and \re{F1-res} hold for arbitrary number of nodes $L$. As expected, the functions $F_L^{(0)}$ and $F_L^{(1)}$ vanish for $L=1$. 
For $L=2$ they are in agreement with the expressions derived in \cite{Beccaria:2023kbl}. 

The relations  \re{FL-free} and \re{F1-res} were derived at strong coupling and fixed $L$ and, therefore, they are valid for $\sqrt\lambda > L$.  
We observe that at large $L$  all terms in the strong coupling expansion \re{FL-free} and \re{F1-res} except the $O(1/g)$ term in \re{FL-free} scale linearly with $L$.  The asymptotic behaviour of the free energy in the opposite limit, for $L \gg \sqrt\lambda \gg 1$ is discussed in the next section.
 
\section{Free energy in the long quiver  limit}\label{sect:large} 

In this section, we examine asymptotic behaviour of the free energy \re{F-exp} in the limit of large number of nodes $L$ at weak and strong coupling.

\subsection*{Weak coupling}

At weak coupling, the dependence of the free energy \re{F-exp} on the number of nodes $L$ can be obtained from \re{F-weak} and \re{F1-weak}. The expansion coefficients in these relations depend on $L$ through the function $\sigma_{2n}(L)$ given by \re{sigma-L}.  

For instance, the weak coupling expansion of the leading nonplanar correction $F^{(0)}_L$ takes the general form (see \re{F-weak})
\begin{align}\label{weak-gen}
F^{(0)}_L = \sum_{n \ge 1}  
\sum_{k_1,\dots,k_n\ge 2}  c_{k_1\dots k_n}\lr{\lambda\over 16\pi^2}^{k_1+\ldots+k_n}\sigma_{2n}(L)\, \zeta_{2k_1-1} \dots \zeta_{2k_n-1}\,,
\end{align}
where $c_{k_1\dots k_n}$ are rational numbers. According to \re{sigma-L}, the function $\sigma_{2n}(L)$ is linear in $L$ for $n<L$ but this behaviour changes starting from $n=L$. \footnote{The relation \re{weak-gen} holds for $L\ge 2$. For $L=2$ it follows from \re{sigma} that $\sigma_{2n}(2)=1$.
In the $L\to \infty$ limit, $\sigma_{2n}(L)$ is replaced with the expression on the second line in \re{sigma-L} leading to  \re{lin-weak}.
}
The function $\sigma_{2n}(L)$ first appears in the expansion \re{weak-gen} at $n=L$ and $k_1=\dots=k_L=2$.
 As a result, at weak coupling, the free energy has the asymptotic behaviour \re{density} only up to order $O(\lambda^{2L})$
\begin{align}
\Delta F_{\QQ} =  L\, \varepsilon(\lambda,1/N) + O(\lambda^{2L})\,.
\end{align}
We apply \re{F-weak} and \re{F1-weak} to obtain the energy density at weak coupling as
\begin{align}\notag\label{e-weak}
\varepsilon(\lambda,1/N) {}& = \frac{3\zeta(3)}{128\pi^{4}}\,\l^{2}-\frac{15\zeta(5)}{1024\pi^{6}}\,\l^{3}+ O(\l^4)
\\
{}& + {1\over N^2} \lr{-\frac{3   \zeta (3)}{128 \pi ^4} \lambda ^2+\frac{25  \zeta (5)}{1024 \pi
   ^6} \lambda ^3 + O(\l^4)} + O(1/N^4)\,.
\end{align}
The breaking of the scaling behaviour \re{density} at order $O(\lambda^{2L})$ is not surprising and has the following interpretation in terms of wrapping, or finite size, effects in the lattice model \re{Z-repr}. 

The partition function \re{Z-repr} describes the propagation of excitations across the lattice with $L$ sites.
Due to the form of the interaction potential \re{Sint}, excitations at site $I$ can jump to the nearest sites $I\pm 1$. At weak coupling, this produces the $O(\lambda^2)$ contribution to the free energy. The propagation of the excitations across $n$ consecutive sites on the lattice generates correction to the free energy of order $O(\lambda^{2n})$. For $n<L$ the propagation range is smaller than 
the circumference of the circle and the corrections to the free energy due to finite size of the system are expected to be small at large $L$. Starting from $n=L$ the excitations can wrap around the lattice and the finite size effects become important. 
This explains why the $L-$dependence of the free energy gets modified at order $O(\lambda^{2L})$.

\subsection*{Strong coupling}
 
At strong coupling, the free energy $\Delta F_{\QQ}$ has different dependence on the number of nodes $L$ depending on the ratio $L/\sqrt\lambda$. 

For $L/\sqrt\lambda\ll 1$, it follows from \re{FL-free} that  the free energy receives $O(L\log L/\sqrt\lambda)$ corrections that do not respect the scaling behaviour \re{density}. Moreover, examining high order corrections to \re{FL-free}, we found that the coefficients of $1/\lambda^{n/2}$ (with $n\ge 2$)  are given by polynomials in $L$ of degree $n$. 
Retaining the terms with the maximal power of $L$  to each order in $1/\sqrt\lambda$, we get
\begin{align}\label{FL-large L} 
 {}& F_L^{(0)}= L \left[  {1\over 6} \sqrt\lambda -  \log {\sqrt\lambda\over 4\pi}  + \lr{\tfrac14-6\log\mathsf A}  -\frac{\log L}{4\sqrt\lambda} \right]-{\sqrt\lambda\over 6L}+ f\lr{L\over \sqrt\lambda}+\dots \,,
\end{align}
where dots denote terms suppressed by powers of $1/L$. As compared with the strong coupling expansion \re{FL-free}, we replaced in \re{FL-large L} the coefficients by their leading behaviour at large $L$ and added the subleading corrections in $L/\sqrt\lambda$. 

These corrections are described by the function $ f(l)$ with $l=L/\sqrt\lambda$. It is given by~\footnote{To obtain this relation, we used the strong coupling expansion of \re{F-ell-exp} obtained in \cite{Belitsky:2020qir} up to order $O(1/\lambda^{15})$.}
\begin{align}\notag\label{f-bad}
f(l) ={}&-\frac{1}{8 }l^2\zeta(2)-\frac{19}{192 }l^3\zeta(3)-\frac{7}{64 }l^4\zeta(4)-\frac{413}{2560 }l^5\zeta(5)-\frac{59}{192
   }l^6\zeta(6)
\\[2mm]
 {}&  -\frac{83875}{114688} l^7\zeta(7)-\frac{4315}{2048 }l^8\zeta(8)-\frac{8433167}{1179648
   }l^9\zeta(9) -\frac{286081}{10240}l^{10} \zeta(10)+\dots\,,
\end{align}
where dots denote corrections suppressed by $1/g$ as well as exponentially small, nonperturbative corrections. One can show following \cite{Beccaria:2022ypy} that the leading nonperturbative correction to \re{f-bad} scales as $O(e^{-2\sqrt\lambda/L})$.

The relation \re{f-bad} is well-defined for $l<1$, or equivalently $1\ll L< \sqrt\lambda$.  
Substituting \re{f-bad} into \re{FL-large L} we find that the free energy does not satisfy \re{density} and has a complicated dependence on the number of nodes $L$. Namely, $F_L^{(0)}$ contains logarithmically enhanced term $-L \log L/(4\sqrt\lambda)$ and a nontrivial function  $ f(L/\sqrt\lambda)$. 

\begin{figure}[t!]
\begin{center}
\includegraphics[width=0.65\textwidth]{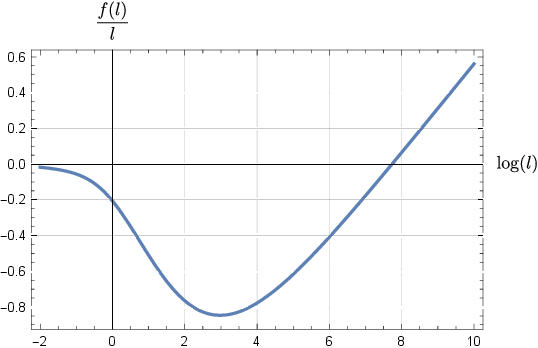}
\caption{Graphical representation of the function $f(l)/l$ defined in  \re{f-resum}. At small  and large $l$ its behaviour is described by \re{f-bad} and \re{f-as}, respectively.}
\label{fig:plot}
\end{center}
\end{figure}%

To find the free energy for $1\ll  \sqrt\lambda \ll L$ using the strong coupling expansion \re{FL-large L}, one has to  perform a resummation of the series \re{f-bad}. This is done in Appendix~\ref{appB}. 
As we show there, the function $f(l)$ admits a closed form representation 
\begin{align}\label{f-resum}
f(l) =  \sum_{\alpha=1}^\infty \left[\log \left(I_0\left(\frac{\alpha}{l}\right) I_1\left(\frac{\alpha}{l}\right)\right)-\frac{2 \alpha  }{l}+\log
   \left(\frac{2 \pi\alpha}{l}\right)+\frac{l}{4 \alpha}\right],
\end{align}
where $I_0(\alpha/l)$ and $I_1(\alpha/l)$ are the modified Bessel functions (not to be confused with the functions $I_n(\chi)$ defined in \re{In}). 
 The last three terms inside the brackets ensure that the sum converges at large $\alpha$. 
It is straightforward to verify that the expansion of \re{f-resum} at small $l$ reproduces the relation \re{f-bad}.
To see this, it is sufficient to replace  the Bessel functions in \re{f-resum} by their asymptotic behaviour at infinity and took into account that $\sum_{\alpha\ge 1} 1/\alpha^n =\zeta(n)$. 
Notice that the small $l$ expansion of \re{f-resum} also contains exponentially small $O(e^{-2/l})$ corrections. 

The dependence of the function \re{f-resum} on $\log l$ is shown in Figure~\ref{fig:plot}. We observe that $f(l)/l$ grows linearly with $\log l$. Indeed, for $l\gg 1$ the dominant contribution to the sum in \re{f-resum} comes from $\alpha \ll l$. Keeping the last, leading term inside the brackets in \re{f-resum} we get for $l \gg 1$
\begin{align}\label{f-as}
f(l) \sim \sum_{\alpha=1}^{c' l} \frac{l}{4 \alpha} = \frac{l}{4}
\lr{ \log (4\pi l)-c}+O(l^0)\,,
\end{align}
where $c=-\gamma - \log (c'/(4\pi))=9.714744$.~\footnote{The constant $c$ was determined by applying the Euler-Maclaurin summation formula to
 \re{f-resum} at large $l$.}

Substituting \re{f-as} into \re{FL-large L} we observe that  troublesome $O(L\log L/\sqrt\lambda)$ term on the right-hand side of \re{FL-large L} cancels against the analogous term coming from \re{f-as}. As a result, the free energy 
takes the form \re{log-l} and exhibits 
the expected scaling behaviour \re{density}. The corresponding expression for the energy density  \re{density} looks as 
\begin{align}\label{eps}
\varepsilon(\lambda,1/N) = {1\over 6} \sqrt\lambda - \log {\sqrt\lambda\over 4\pi} + \lr{\tfrac14-6\log\mathsf A}  -\frac{1}{4\sqrt\lambda} \lr{ \log {\sqrt\lambda\over 4\pi}+c} +O(1/N^2)\,.
\end{align}
We would like to emphasize that this relation holds for $1\ll  \sqrt\lambda \ll L$.

\section{Circular Wilson loop} \label{sect5} 

In this section, we compute the expectation value of the circular half-BPS Wilson loop in the $\QQ$ model. It is defined as
\begin{align}\label{W}
W^{\QQ} = \VEV{\tr P \exp\lr{g_{\rm YM} \oint ds \left[i A^\mu _{I}(x) \dot x_\mu(s) +{1\over \sqrt 2}(\varphi_I(x)+\varphi_I^*(x))\right]}}\,,
\end{align}
where the gauge field $A_I^\mu$ and complex scalar field $\varphi_I(x)$ are integrated along a circle of unit radius. These fields
belong to the $SU(N)$ $\mathcal N=2$ vector multiplet at node $I$. The Wilson loop \re{W} does not depend on the choice of the node. In what follows we choose $I=1$.

For $L=2$ the Wilson loop \re{W} was computed in the $\mathsf Q_2$ model in \cite{Beccaria:2023kbl}. The subsequent analysis of \re{W} goes along the same lines as in that paper. The localization gives the representation of \re{W} as the ratio of expectation values
\begin{align}\label{W-vev}
W^{\QQ}= {\VEV{\tr\lr{e^{\sqrt{\lambda\over 2N} A_1}} \, e^{\sum_I S_{\rm int}(A_I,A_{I+1})}}_0 \over \VEV{e^{\sum_I S_{\rm int}(A_I,A_{I+1})}}_0}\,,
\end{align}
where the interaction potential $S_{\rm int}(A_I,A_{I+1})$ is given by \re{Sint}. As before, the subscript `$0$' indicates that the average is evaluated with the Gaussian measure for the matrices $A_I$ with $I=1,\dots, L$. 

We recall that the interaction potential $S_{\rm int}(A_I,A_{I+1})$ describes the coupling of the matrices in the adjacent nodes. Neglecting this potential in \re{W-vev} we recover the expectation value of the circular Wilson loop in $\mathcal N=4$ SYM theory
\begin{align}\label{W4}
W^{\mathcal N=4} = \VEV{\tr\lr{e^{\sqrt{\lambda\over 2N} A_1}}  }_0  = {2 N\over \sqrt\lambda} I_1(\sqrt\lambda) + O(1/N)\,.
\end{align}
The difference between the Wilson loop in the two models $W^{\QQ}-W^{\mathcal N=4}$ arises due to the interaction between the matrices in the different nodes described by \re{Sint}. Due to a peculiar form of the potential \re{Sint}, the interaction does not affect the leading $O(N)$ contribution to \re{W-vev}. As a consequence, the Wilson loops $W^{\QQ}$ and $W^{\mathcal N=4}$ coincide in the planar limit. 
In a close analogy with the free energy \re{free}, this suggests to define the difference function
\begin{align}\label{Del-W}
\Delta W^{\QQ}= W^{\QQ}-W^{\mathcal N=4}\,.
\end{align}
At large $N$ its expansion starts at order $O(1/N)$. 

Expanding \re{W-vev} in powers of $A_1$ we can express $W^{\QQ}$ as an infinite sum over expectation value of single traces \re{O}
\begin{align}\label{W-sum}
W^{\QQ} = N + {1\over Z_{\QQ}} \sum_{n\ge 1} {1\over (2n)!} \lr{\lambda\over 2}^n 
\VEV{\mathcal O_{2n}(A_1) \, e^{\sum_I S_{\rm int}(A_I,A_{I+1})}}_0\,,
\end{align}
where $Z_{\QQ}=\VEV{e^{\sum_I S_{\rm int}(A_I,A_{I+1})}}_0$ is the partition function of the model defined in \re{Z-repr}. In the absence of the interaction, for $S_{\rm int}(A_I,A_{I+1})=0$, the expectation value $\vev{\mathcal O_{2n}(A_1)}$ can be obtained by differentiating the 
generating function \re{gf} with respect to the source, $\partial Z(J_1)/\partial J_{1,2n}$. According to \re{int-J}, the interaction term can be generated by averaging this derivative over the source fields with the measure  \re{Z-Feynman}. 

In application to \re{W-sum} this leads to the following representation
\begin{align}\label{W-G}
W^{\QQ} = N +\sum_{n\ge 1} {1\over (2n)!} \lr{\lambda\over 2}^n \left[ N G_{2n} +{1\over 2N} \vev{J_{1,i} J_{1,j}} G_{2n,ij}  +O(1/N^3) \right],  
\end{align}
where $G_{2n,ij\dots}$ are connected correlation functions in the Gaussian matrix model \re{G} and $\vev{J_{1,i} J_{1,j}}$ denotes an average with respect to the measure \re{Z-Feynman}. At large $N$, we use \re{sour} and \re{prop} to get
\begin{align}
\vev{J_{1,i} J_{1,j}} =  {1\over L} \sum_{\alpha,\alpha' =1}^{L-1} e^{i(p_\alpha+p_{\alpha'})}\vev{\widetilde J_{i}(p_\alpha)\widetilde J_{j}(p_{\alpha'})}=  {1\over L} \sum_{\alpha=1}^{L-1} X_{ij}(p_\alpha)\,,
\end{align}
where the semi-infinite matrix $X_{ij}(p_\alpha)$ is defined in \re{X}. Notice that it is proportional to the interaction matrix $C$.

In the special case of $\mathcal N=4$ SYM, for $S_{\rm int}(A_I,A_{I+1})=0$, only the first term inside the brackets in \re{W-G} survives and $W^{\mathcal N=4}$ is given by the sum over $G_{2n}$. Together with \re{Del-W} this leads to
\begin{align}\label{WL}
\Delta W^{\QQ}={1\over 2NL} \sum_{n\ge 1} {1\over (2n)!} \lr{\lambda\over 2}^n   \sum_{\alpha=1}^{L-1} X_{ij}(p_\alpha) G_{2n,ij}   + O(1/N^3)\,.
\end{align}
For $L=2$ the sum over $\alpha$ contains only one term with the quasimomentum $p=\pi$. In this case, replacing the three-point function $G_{2n,ij}$ with its leading large $N$ expression, one gets \ci{Beccaria:2023kbl}
\begin{align}\label{W2}
\Delta W^{\mathsf Q_2}={1\over 16N} \rho \left[w_{00}^+(\chi) + w_{00}^-(\chi)\right]  + O(1/N^3)\,,
\end{align}
where the matrix elements $w_{00}^\pm (\chi)$ are given by \re{w-vev} for $L=2$ and the notation was introduced for
\begin{align}
\rho=\sum_{n\ge 1} {1\over (2n)!} \lr{\lambda\over 2}^n  n(n+1) G_{2n} = \frac{\lambda}{4N} W^{\mathcal N=4} + O(1/N^2)\,.
\end{align}
Here in the second relation we used \re{W4}.

Going from \re{W2} to \re{WL}, it is sufficient to multiply \re{W2} by the factor of $2/L$ and replace the symbol inside the matrix elements $w_{00}^\pm (\chi)$ according to \re{chi-sub}. This leads to
\begin{align}\label{del-W1}
\Delta W^{\QQ}=\frac{\lambda}{32 L N^2}  W^{\mathcal N=4}\left[\vvev{w_{00}^+} + \vvev{w_{00}^-}\right]  + O(1/N^3)\,.
\end{align}
where $\vvev{w_{00}^\pm}$ are given by \re{w-vev}. Taking into account \re{F0-sum} and \re{w00} we observe that the sum of matrix elements $\vvev{w_{00}^+} + \vvev{w_{00}^-}$ is equal to a derivative of the free energy $(-4g\partial_g F_L^{(0)})$. As a consequence, the relation \re{del-W1} can be written as
\begin{align}
\Delta W^{\QQ}=-\frac{\lambda^2 }{4 L N^2}  W^{\mathcal N=4} \partial_\lambda F_L^{(0)} + O(1/N^3)\,.
\end{align}
Being combined with \re{Del-W} it leads to the following remarkably simple relation between the ratio of the Wilson loops in the two theories and the  free energy
\begin{align}\label{rat}
{W^{\QQ}\over W^{\mathcal N=4}} = 1 - \frac{1}{4 L N^2}\lambda^2\partial_\lambda F_L^{(0)} + O(1/N^4)\,.
\end{align}
We would like to emphasize that this relation is valid for an arbitrary coupling $\lambda$.  As expected, the Wilson loops $W^{\QQ}$ and $W^{\mathcal N=4}$ coincide in the leading large $N$ limit. It is interesting to note that the nonplanar $O(1/N^2)$ correction to \re{rat} is negative for arbitrary coupling. \footnote{Notice that $(-\partial_\lambda \log Z_{\QQ})$ is given by an expectation value of a positive definite function, as follows from \re{Z-Q} and \re{S-Q2}.}

\begin{figure}[t!]
\begin{center}
\includegraphics[width=0.5\textwidth]{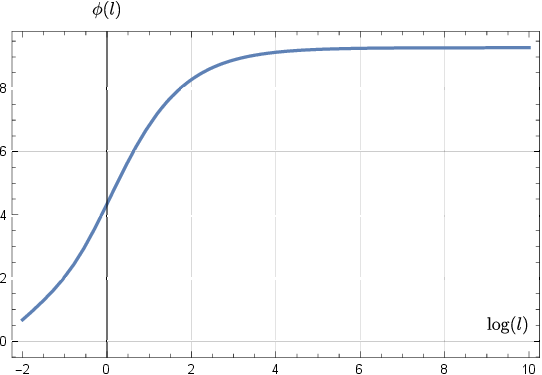}
\caption{Graphical representation of the function $\phi(l)$ defined in \re{phi1}.}
\label{fig:W}
\end{center}
\end{figure}%

Let us examine the ratio of Wilson loops \re{rat} at large $L$. At weak coupling we apply \re{density} and \re{e-weak} to find  
\begin{align}\notag\label{WoverW1}
{W^{\QQ}\over W^{\mathcal N=4}} {}& = 1 - \frac{1}{4 N^2}\lambda^2\partial_\lambda \varepsilon  + O(1/N^4)
\\
{}& =1 - \frac{1}{N^2}\left[\frac{3 \lambda ^3 \zeta (3)}{256 \pi ^4}-\frac{45 \lambda ^4 \zeta (5)}{4096 \pi ^6} +O(\l^5)\right] +O(1/N^4)
 \,.
\end{align}
Note that the ratio is independent on $L$ and it approaches a finite value for $L\to\infty$.

At strong coupling, we use \re{FL-large L} to obtain from \re{rat}
\begin{align}\label{WoverW}
{W^{\QQ}\over W^{\mathcal N=4}} {}& =1 -\frac{1}{N^2}\left[\frac{\lambda ^{3/2}}{48}-\frac{\lambda }{8}+\frac{1}{32} \sqrt{\lambda } \left(\log {\sqrt\lambda\over 4\pi}+\phi\lr{L\over \sqrt\lambda}\right)\right]+O(1/N^4)+O(1/L)\,.
\end{align}
where the notation was introduced for 
\begin{align}\label{phi1}
\phi(l)=\log (4\pi l) - 4{df(l)\over dl}\,.
\end{align}
The dependence of this function on $\log l$ is shown in Figure~\ref{fig:W}. It grows at small $l$ as $\phi(l)=\log (4\pi l)+ \zeta(2) \, l+ O(l^2)$ and for  $l\to\infty$ it approaches a constant $\phi(l)=c-1$ where $c$ is defined in \re{f-as}.

The relation \re{WoverW} holds for large $\sqrt\lambda$ and $L$. 
Similar to the free energy, the expression on the right-hand side of \re{WoverW} has different dependence on the number of nodes depending on how $L$ compares with $\sqrt\lambda$. For $L/\sqrt\lambda \ll 1$ and $L/\sqrt\lambda \gg 1$, the ratio \re{WoverW} is given by \re{W-lim1} and \re{W-lim}, respectively.  Like the energy density \re{eps}, it ceases to depend on the number of nodes for $1\ll \sqrt\lambda\ll L$. This property requires an explanation.

\section*{Acknowledgments}

We are very grateful to Arkady Tseytlin for collaboration on subjects closely related to this work.
We would like to thank Sergey Frolov, Zohar Komargodski, Paul Krapivsky and Pierfrancesco Urbani for very useful discussions.  One of us (G.K.) is grateful to the Hamilton Mathematics Institute at Trinity College Dublin for kind hospitality extended to him where essential part of this work was done.
  
\appendix

\section{Strong coupling expansion}\label{app}

In this appendix, we present some details of derivation of the strong coupling expansion of the free energy \re{FL-free}. The expression for $F_L^{(0)}$
  can be obtained from \re{F0-sum} by replacing the function $\mathcal F_{\ell}(s_\alpha \chi)$ with its asymptotic expression 
 \re{F-ell-exp}.  
This leads to
\begin{align}\label{F0-aux} 
F_L^{(0)} = -2 g \vvev{I_0} - \log g+ \frac12 \vvev{B_{\ell=1}+B_{\ell=2}}
- {1\over 16g} \vvev{I_1}- {1\over 64 g^2} \vvev{I_1^2} + O(1/g^3)  \,,
\end{align}
where the expansion parameter $g$ is defined in \re{g} and $\vvev{f}$ denotes the sum of the functions $f(s_\alpha\chi)$ evaluated for possible values of the quasimomentum
\begin{align}\label{vvev}
\vvev{f} =  \sum_{\alpha=1}^{L-1} f(s_\alpha\chi)\,,\qqqquad s_\alpha=\sin^2\left(\pi \alpha\over L\right)\,.
\end{align}
Taking into account \re{I0} and going through the calculation, we get
\begin{align}\notag
& \vvev{I_0} = -\frac{\pi \left(L^2-1\right)}{3 L}\,,\qquad 
\\[1.5mm]
{}&\notag
\vvev{I_1}={L \log L \over \pi  }\,, 
\\
& \vvev{I_1^2} = \frac{L(L-1) }{\pi ^2}
\left[ \sum _{\alpha=1}^{L-1} \log ^2\left(\sin \left(\frac{\pi 
   \alpha}{L}\right)\right)-L\log ^2(2)+\log^2 (2L) \right].
\end{align}
Higher order corrections to \re{F0-aux} involve sums of the form $\vvev{I_1^{n_1} I_2^{n_2}\dots}$ which can be computed in the similar manner.

\subsection*{Widom-Dyson constant}

The constant $O(g^0)$ term in \re{F0-aux} involves the Widom-Dyson constant $B_\ell(s_\alpha \chi)$. It is given by
\cite{Belitsky:2020qrm,Belitsky:2020qir}
\begin{align}\notag\label{B-const}
{}& B_\ell(s_\alpha \chi) =W(\alpha/L) -\frac12 \log(2\pi) -{\ell\over 2} \log (4s_\alpha) + \log\Gamma(\ell)\,,
\\\notag
{}& 
W(\alpha/L) = \frac12 \int_0^\infty dk \left[k \lr{\widetilde\psi(k) }^2-{1-e^{-k}\over k} \right],
\\
{}& \widetilde\psi(k) = \int_0^\infty {dx\over\pi} \cos(kx) \log(1-s_\alpha   \chi(x))\,,
\end{align} 
where $s_\alpha=\sin^2(\pi \alpha/L)$ and $ \chi(x)=-1/\sinh^2(x/2)$.
Performing integration we find
\begin{align} \notag
\widetilde\psi(k) {}&=\frac{2  \sinh \left(\pi k \ft{\alpha}{L}\right) \sinh \left(\pi  k(1-\ft{\alpha}{L})\right)}{k \sinh (\pi k)}\,,
\\
W(r) {}&=\frac{1}{2} \log \left[\frac{4 \pi ^2 (2 r-1) [G(1-r) G(r)]^4 \cot (\pi  r)}{G(1-2 r)
   G(2 r-1)}\right],
\end{align}
where $r=\alpha/L$ and $G(x)$ is Barnes function. It is easy to check that $W(r)=W(1-r)$. In this way, we obtain the $O(g^0)$ term in \re{F0-aux} as
\begin{align}\label{app:C} 
&C_L \equiv \vvev{B_{\ell=1}+B_{\ell=2}}=2\sum_{\alpha=1}^{L-1} W(\alpha/L)-(L-1)\log(2\pi)-3 \log L\,.
\end{align}
The explicit expressions for $C_L$ for the few lowest values of $L$ are given by \re{C-vals}. 

Combining together the above relations we arrive at \re{FL-free}. 

\subsection*{Matrix elements}

The nonplanar correction \re{F1-QL} to the free energy $F_L^{(1)}$ depends on the matrix elements of the resolvent of the Bessel operator \re{w}.

At weak coupling, we can expand \re{w} in powers of the Bessel operator \re{K-op} to get
\begin{align}\label{w-weak}
w_{nm}(s_\alpha \chi) {}&= s_\alpha \vev{\phi_n|\bm{\chi}   |\phi_m}+   s_\alpha^2 \vev{\phi_n|\bm{\chi} {  \bm K_\ell} |\phi_m} + \dots\,.
\end{align}
The matrix elements on the right-hand side are given by
\begin{align}\notag\label{app:vev}
\vev{\phi_n|\bm{\chi}   |\phi_m} {}&= \int_0^\infty dx\, \phi_n(x) \phi_m(x) \chi\lr{\sqrt x\over 2g} 
\\
{}& = (2g)^2  \int_0^\infty dx\, \phi_n((2g)^2x) \phi_m((2g)^2x) \chi\lr{\sqrt x} 
\end{align}
where in the second relation we changed the integration variable $x\to (2g)^2 x$. Replacing the function $\phi_n(x)$ with its expressions \re{phi}, one can expand \re{app:vev} in powers of $g^2$. In a similar manner, the second term on the right-hand side of \re{w-weak} can be evaluated as
\begin{align}
\vev{\phi_n|\bm{\chi} {\bm K_\ell} |\phi_m} {}&= \sum_{i\ge 1} \vev{\phi_n|\bm{\chi}   |\psi_i} \vev{\psi_i|\bm{\chi}   |\phi_m}\,,
\end{align}
where we took into account that $ {\bm K_\ell}=\sum_{i\ge 1} \ket{\psi_i}\bra{\psi_i}\bm{\chi}$. Here the matrix elements $\vev{\phi_n|\bm{\chi}|\psi_i}$ are given by \re{app:vev} with
$\phi_m$ replaced by the function $\psi_i$ defined in \re{psi}. The resulting expressions for the matrix elements
 at weak coupling are
\begin{align}\label{m.e.}\notag
w_{00}^+(s_\alpha\chi) &=4 g^4 q_1-4 g^6 q_2+g^8 \left(2 q_1^2+\frac{5 q_3}{3}\right)+\dots,
\\\notag
w_{00}^-(s_\alpha\chi) &=g^6 q_2-\frac{2 g^8 q_3}{3}+\frac{7 g^{10} q_4}{36}+\dots,
\\\notag
w_{01}^+(s_\alpha\chi) &=2 g^4 q_1-4 g^6 q_2+g^8 \left(q_1^2+\frac{5 q_3}{2}\right)+\dots ,
\\\notag
w_{01}^-(s_\alpha\chi) &=g^6 q_2-g^8 q_3+\frac{7 g^{10} q_4}{18}+\dots,
\\\notag
w_{11}^+(s_\alpha\chi) &=g^4 q_1-3 g^6 q_2+g^8 \left(\frac{q_1^2}{2}+\frac{37 q_3}{12}\right)+\dots,
\\\notag
w_{11}^-(s_\alpha\chi) &=g^6 q_2-\frac{4 g^8 q_3}{3}+\frac{25 g^{10} q_4}{36}+\dots,
\\\notag
w_{02}^+(s_\alpha\chi) &=  g^4 q_1-5 g^6 q_2+g^8 \left(\frac{q_1^2}{2}+\frac{53 q_3}{12}\right) +\dots,
\\
w_{02}^-(s_\alpha\chi) &= g^6 q_2-\frac{5 g^8 q_3}{3}+\frac{31 g^{10} q_4}{36}+\dots,
\end{align}
where the superscripts `$+$' and `$-$' correspond to $\ell=1$ and $\ell=2$, respectively. 
Here the notation was introduced for
\begin{align}
q_n= 2s_\alpha \int_0^\infty dx\, x^{2n+1}\chi(x) = -8s_\alpha (2n+1)! \zeta(2n+1)\,,
\end{align}
where $\chi(x)$ is given by \re{chi}. Substituting the relations \re{m.e.} into \re{F1-QL} and \re{w-vev}, we arrive at \re{F1-weak}. It is straightforward to verify that the matrix elements \re{m.e.} satisfy the functional relations \re{fun-rel}. 

At strong coupling, we use the expressions for the matrix elements \re{w-nm} derived in \cite{Beccaria:2023kbl}
\begin{align}\notag\label{w-expr}
w_{00}(\chi) &=4 g I_0 +(2 \ell -1)-\frac{ (2 \ell -3) (2 \ell -1) I_1 }{8 g}-\frac{(2 \ell -3) (2 \ell -1)
   I_1^2 {} }{16 g^2} + \dots
\\\notag
w_{01}(\chi) &=-2 g^2 I_0^2-2 g  \ell  I_0 +\frac{1}{8} (2 \ell -1) \left((2 \ell -3) I_0
   I_1-2 \ell -3\right)
\\ \notag&   
   +\frac{(2 \ell -3) (2 \ell -1)  }{16 g}I_1 \left(I_0 I_1+\ell
   +1\right)+\dots
\\
\notag
w_{02}(\chi) {}&=
\frac{2}{3} g^3 \left(I_0^3 -2 I_{-1} \right)+\frac{1}{2} g^2 (2 \ell -1) I^2_0 
   +g \left(\ell ^2
   I_0 -\frac{1}{16} (2 \ell -3) (2 \ell -1)I^2_0  I_1 \right)
\\   \notag
{}&   
 -\frac{1}{32} (2 \ell -1) \left(I_0^2 I_1^2 (2 \ell -3)+I_0 I_1 (2
   \ell -3) (2 \ell +1)-4 \ell ^2-4 \ell -3\right)+\dots
\\   \notag
w_{11}(\chi){}& = \frac{4}{3} g^3 \left(I_0^3{} +I_{-1}\right)+2 g^2 \ell  I^2_0{} +\frac{1}{8} g I_0
   \left(-(2 \ell -3) (2 \ell -1) I_0 I_1+4 \ell ^2+8 \ell -3\right)
\\ &  
     -\frac{1}{16} (2 \ell -1) \left(I_0^2 I_1^2 (2 \ell -3)+2 I_0 I_1 (\ell +1) (2 \ell -3)-4
   \ell -3\right)+\dots 
\end{align}
These relations involve the functions $I_n=I_n(\chi)$. For $n\ge 0$ they are given by \re{In}. For $n<0$ we have instead
\begin{align}
I_n(\chi) =   \int_0^\infty {dx\over\pi} x^{1-2n}\partial_x\log\left(1-\chi(x)\right).
\end{align}
To obtain the strong coupling expansion of the matrix elements \re{w}, it is sufficient to replace $I_n(\chi)$ with $I_n(s_\alpha\chi)$ in \re{w-expr} and apply \re{I0} together with
\begin{align}
I_{-1}(s_\alpha\chi) = - 4\pi^3 {(\alpha(L-\alpha))^2\over L^4}\,.
\end{align} 

\subsection*{The contribution of diagram in Figure~\ref{fig}(c)}

The relations \re{w-expr} allow us to compute expressions on the first two lines of \re{F1-QL}. The last line of \re{F1-QL} describes the contribution of the diagram shown in Figure~\ref{fig}(c). It  contains a sum over the quasimomenta propagating inside the loops. 
It is convenient to rewrite this sum as
\begin{align}\label{d-sum}
 -{1\over 768 L} 
  \sum_{p_1,p_2,p_3} \delta_{p_1+p_2+p_3}w^+_{0,0}(p_1)
\Big(w^+_{0,0}(p_2)w^+_{0,0}(p_3)
 + 3w^-_{0,0}(p_2)w^-_{0,0}(p_3)\Big)\,,
\end{align}
where $w^\pm_{0,0}(p_i) \equiv w^\pm_{0,0}(s_{\alpha_i} \chi)$ and
$p_i=2\pi \alpha_i/L$ (with $1\le \alpha_i\le L-1$). 

The $\delta-$function in \re{d-sum} imposes the momentum conservation \re{delta}.
Replacing 
\begin{align}\label{d-r}
\delta_{p_{1}+p_{2}+p_{3}} = {1\over L}\sum_{n=1}^L e^{-in (p_{1}+p_{2}+p_{3})}\,,  \qqqquad
 \widetilde w_n^\pm= \sum_{\alpha=1}^{L-1} e^{-in p_\alpha} w^\pm_{0,0}(s_{\alpha_i}\chi)
\end{align} 
we can rewrite \re{d-sum} as 
\begin{align}\label{app:w}
 -{1\over 768 L^2}\sum_{n=1}^L \left[ (\widetilde w_n^+)^3 + 3 \widetilde w_n^+(\widetilde w_n^-)^2 \right]
 \,.
\end{align}
This sum can be thought of as a discretized  version of the integral $\int dx\, [D (x)]^3$ involving a power of the propagator. 
By definition, $\widetilde w_n^+$ and $\widetilde w_n^-$ are given by the function
\begin{align}
\widetilde w_n {}& = \sum_{\alpha=1}^{L-1} e^{-in p_\alpha} w _{00}(s_\alpha\chi)
\end{align}
evaluated for $\ell=1$ and $\ell=2$, respectively.  

Taking into account the first relation in \re{w-expr}, we get
\begin{align}\notag
\widetilde w_n {}& =\sum_{\alpha=1}^{L-1} e^{-in p_\alpha} \left[ 4 g I_0(s_\alpha\chi)+(2 \ell -1) 
-\frac{ (2 \ell -3) (2 \ell -1)  }{8 g}I_1(s_\alpha\chi)+ O(1/g^2)\right],
\end{align}
where $ I_0(s_\alpha\chi)$  and $ I_1(s_\alpha\chi)$ are given by \re{I0}.
Notice that $\sum_{n=1}^L \widetilde w_n=0$.
Explicit expression for $\widetilde w_n$ is different for $1\le n\le L-1$ and $n=L$. In the former case we have
\begin{align}\label{app:w1}
\widetilde w_n {}&={4\pi g\over  L \sin^2(n\pi/L)} -(2\ell-1) -\frac{ (2 \ell -3) (2 \ell -1)  }{8 g}\sum_{\alpha=1}^{L-1} e^{-in p_\alpha}  I_1(s_\alpha\chi)+ O(1/g^2)\,.
\end{align}
For $n=L$ we have instead
\begin{align} \label{app:w2}
\widetilde w_L {}&=-\frac{4 \pi  g (L-1) (L+1)}{3 L}+(2\ell-1)(L-1)-\frac{ (2 \ell -3) (2 \ell -1)  }{8 g}{L\over \pi}\log L + O(1/g^2)\,.
\end{align} 
These expressions are valid for arbitrary $L\ge 1$.

The functions $\widetilde w_n^+$ and $\widetilde w_n^-$ are given by \re{app:w1} and \re{app:w2} for $\ell=1$ and $\ell=2$, respectively. 
Their substitution into  \re{app:w} leads to
\begin{align}\notag\label{DF1} 
 {}&(g\pi)^3 \left(\frac{52  }{945 L^5}-\frac{1}{45 L^3}+\frac{11  L}{945}-\frac{2
  }{45 L}\right) 
  +(g\pi)^2 \left(-\frac{1}{15 L^4}-\frac{1}{18 L^3}-\frac{  L}{18}+\frac{1}{9 L}+\frac{1}{15}\right)
 \\&
 +(g\pi)\left(\frac{\vvev{I_0 I_1}}{48 L^2}-\frac{\vvev{I_0^2 I_1}}{192 \pi
   }-\frac{\log L}{240 L^3}+\frac{1}{6 L^2}+\frac{L}{12}-\frac{1}{12 L}+\frac{1}{240} L
   \log (L)-\frac{1}{6} \right)
  +O(g^0)\,,
\end{align}
where $\vvev{I_0^n I_1}$ is defined according to \re{vvev}. 
One can verify that the expression \re{DF1} vanishes at $L=2$ as it should be.

The relation \re{DF1} yields the sum on the last line in \re{F1-QL}. Adding the contribution of the first two lines in \re{F1-QL}, we arrive at the relation \re{F1-res}. 
Notice that terms containing $\log L$ and $\vvev{I_0^2I_1}$ cancel in $F_L^{(1)}$.

\section{Resummation}\label{appB}

In this appendix, we derive the relation \re{f-resum}. We recall that the function $f(L/\sqrt\lambda)$ defines the subleading correction to the free energy \re{FL-large L} in the double scaling limit  
\begin{align}\label{double}
\lambda \gg 1\,,\qqqquad  L\gg 1\,,\qqqquad  L/\sqrt\lambda=\text{fixed}\,.
\end{align}

 \subsection*{Leading behaviour}
 
For arbitrary $L$ the free energy \re{F0-sum} is given by the sum over possible values of the quasimomentum $p_\alpha={2\pi \alpha/L}$. At large $L$, the quasimomentum takes continuous values $0\le p\le  2\pi$. This suggests to rewrite the free energy \re{F0-sum} as an integral over $p$
\begin{align}\label{FL-sum1} 
F_L^{(0)} &=\frac{L}2\sum_{\ell=1,2}  \int_0^{2\pi} dp\, \rho(p)  \mathcal F_\ell(\sin^2(p/2)\chi)\,.
\end{align}
Here we replaced $s_\alpha$ with its expression \re{s} and introduced notation for the density function
\begin{align}\label{rho-norm}
\rho(p) = \frac1{L}\sum_{\alpha=0}^{L-1}\delta(p-2\pi\alpha/L)\,. 
\end{align}
It satisfies the normalization condition $\int_0^{2\pi} dp\,\rho(p)=1$ and admits the large $L$ expansion~\footnote{
This relation follows from the
Euler-Maclaurin summation formula for $\sum^{L-1}_{\alpha=0}\Phi(2\pi\alpha/L)=L \int_0^{2\pi} dp\, \rho(p) \Phi(p)$ where $\Phi(p)$ is a generic test function.} 
\begin{align}\label{rho-exp}
\rho(p) = {1\over 2\pi} -{1\over 2L}\lr{\delta(p-2\pi)-\delta(p)}-{\pi\over 6L^2}\lr{\delta'(p-2\pi)-\delta'(p)}+O(1/L^3)\,,
\end{align}
where prime denote a derivative over $p$. 

At strong coupling, we can replace the function $\mathcal F_\ell(\chi)$ in \re{FL-sum1} by its leading behaviour $\mathcal F_\ell(\chi) = -2g I_0(\chi) +O(g^0)$ (see \re{F-ell-exp}). Taking into account \re{In} and \re{chi}, we obtain the contribution of the first term in \re{rho-exp} to \re{FL-sum1} as
\begin{align}\label{L-term}
- 2L g \int_0^{2\pi} {dp\over 2\pi} \int_0^\infty {dx\over \pi} x \partial_x \log\left(1+{\sin^2(p/2)\over \sinh^2(x/2)}\right) = L {2g \pi\over 3}\,.
\end{align}
Repeating the calculation using the $O(1/L)$ term in \re{rho-exp}, one finds that it yields a vanishing contribution to \re{FL-sum1}.
The contribution of the $O(1/L^2)$ term in \re{rho-exp} to \re{FL-sum1} is 
\footnote{Here the integral over $x$ gives $2|\sin (p/2)|/\sin(p/2)$ and it approaches opposite values $\pm 2$ for $p\to 0$ and $p\to2\pi$.}
\begin{align}\label{FL-pre}
-{g\pi\over 6L}\int_0^{2\pi}dp\,\partial_p \int_0^\infty{dx\over \pi} x \partial_x \log\left(1+{\sin^2(p/2)\over \sinh^2(x/2)}\right) = -{2g\pi\over 3L}\,.
\end{align}
Adding together \re{L-term} and \re{FL-pre}, we reproduce the first term in the expression \re{FL-free} for
$F_L^{(0)}$.

Notice that the integrals \re{L-term} and \re{FL-pre} have different behaviour in the double scaling limit \re{double}. Both integrals receive a dominant contribution from $x\sim |\sin(p/2)|$ but the corresponding values of the quasimomentum are different. The leading contribution to \re{L-term} comes  from the region $0<p<2\pi$ whereas for \re{FL-pre} it only comes from the end-points, $p\to 0$ and $p\to 2\pi$, or equivalently $\sin(p/2)\to 0$.

Subleading corrections to the density function \re{rho-exp} give rise to the function $f$ defined in \re{FL-large L}.
An important observation is that, similar to \re{FL-pre}, it arises from the integration in \re{FL-sum1} over the end-point region $x\sim |\sin(p/2)|$ and $p \to 0 \ \text{(mod $2\pi$)}$.  In application to \re{F-ell-exp} this implies that,  computing the subleading corrections to \re{FL-sum1}, we can replace the symbol $\chi(x)$ in the definition of the function \re{In} with its asymptotic behaviour at small $x$.
According to \re{chi},  the corresponding symbol is
\begin{align}\label{chi-simp}
\chi_0(x) = -{4\over x^2}\,.
\end{align}
 We would like to emphasize that, substituting $\chi(x)$ with $\chi_0(x)$ in \re{FL-sum1}, we only expect to recover the subleading correction to $F_L^{(0)}$ but not the leading one. The latter  correction comes from the integration in \re{L-term} over finite $x$ in which case such substitution is not justified. 

Replacing $\chi(x)$ with $\chi_0(x)$ in \re{FL-sum1} leads to a significant simplification. The main reason is that 
 the function \re{F-ell} can be found in a closed form for the symbol \re{chi-simp} (see Appendix~D in \cite{Beccaria:2023kbl})
\begin{align}\label{F-chi0}
\mathcal F_\ell(\chi_0) =  \log \left( \Gamma(\ell)  {I_{\ell-1}(4g) \over (2g)^{\ell-1}}\right)\,,
\end{align}
where $I_{\ell-1}(4g)$ is a modified Bessel function of the first kind.
To obtain $\mathcal F_{\ell}(s_\alpha \chi_0)=\log\det(1-K_\ell(s_\alpha\chi_0))$ from \re{F-chi0}, we take into account that the semi-infinite matrix $K_\ell(s_\alpha\chi_0)$
defined in \re{K-mat} involves the product $s_\alpha \chi_0\lr{x/(2g)}=-4g^2\sin^2(p_\alpha/2)/x^2$. Because the coupling constant is 
accompanied by $\sin(p_\alpha/2)$, the dependence of $\mathcal F_{\ell}(s_\alpha \chi_0)$ on $s_\alpha$ can be restored by replacing $g\to g \sin(p_\alpha/2)$ on the right-hand side of \re{F-chi0}. In addition, in the end-point region, for $p_\alpha\to 0$, we can replace $\sin(p_\alpha/2)$ with $p_\alpha/2$.  
Applying the transformations outlined above, we get from \re{F0-sum} and \re{F-chi0}
\begin{align}\label{FL-reg}
F_L^{(0)} {}&\sim  \sum_{1\le \alpha\ll L}\log  \lr{  {1\over g  p_\alpha }I_0\lr{2g p_\alpha} I_1\lr{2g p_\alpha} } \,.
\end{align}
Here we took into account that terms with $\alpha \ll L$ and $(L-\alpha)\ll L$ provide the same contribution to the sum. 
 
By construction, the relation \re{FL-reg} captures the contribution to \re{FL-sum1} from the end-point region $\sin(p/2)\to 0$. As explained above, we expect that it will generate the function $f$  in the relation \re{FL-large L}.  Indeed, expanding \re{FL-reg} at large $g$ we obtain
\begin{align}\label{F-as}
F_L^{(0)}\sim \sum_{1\le \alpha\ll L} \lr{4 g p_\alpha -\log \left(4 \pi 
   g^2 p_\alpha^2\right)-\frac{1}{8 g p_\alpha} -\frac{1}{32 g^2 p_\alpha^2}-\frac{19}{1536 g^3 p_\alpha^3}  -\frac{7}{1024 g^4 p_\alpha^4}+\dots},
\end{align}
where $p_\alpha=2\pi\alpha/L$. We observe that the first three terms inside the brackets provide a contribution that diverges in the double scaling limit \re{double}. The contribution of the remaining terms remains finite in this limit. Moreover, replacing 
$\sum_\alpha 1/p_\alpha^n= L^n\zeta(n)/(2\pi)^n$ as $L\to\infty$, we find that it correctly reproduces the expansion \re{f-bad} of the function $f(l)$ with $l=L/(4\pi g)$. 
Subtracting from \re{FL-reg} the contribution of the first three terms in \re{F-as}, we arrive at the following 
relation
\begin{align}\label{delta-F}
f =  \sum_{\alpha=1}^\infty \log \lr{4 \pi  g p_\alpha I_0(2 g p_\alpha) I_1(2 g p_\alpha )e^{\frac{1}{8 g p_\alpha}-4 g p_\alpha }}\,.
\end{align}
It is straightforward to verify that the large $g$ expansion of \re{delta-F} coincides with \re{f-bad} up to exponentially small $O(e^{-8\pi g/L})$ corrections.
Replacing $p_\alpha=2\pi\alpha/L$ and $g=\sqrt\lambda/(4\pi)$ in \re{delta-F}, we finally arrive at \re{f-resum} with $l=L/\sqrt\lambda$.
   
\bibliographystyle{JHEP} 

\providecommand{\href}[2]{#2}\begingroup\raggedright\endgroup

\end{document}